\documentclass[10pt]{aastex}

\usepackage{graphicx,natbib}
\usepackage{amssymb}
\usepackage{epstopdf}

\newcommand{\dpart}[2]{\frac{\partial #1}{\partial #2}}

\newtheorem{theorem}{Theorem}

\shorttitle{PIAA Theoretical}
\shortauthors{Pueyo et al.}

\begin{document}

\title{Design of PIAA coronagraphs over square apertures}
 

\author{Laurent Pueyo  \altaffilmark{1}  \altaffilmark{2} \altaffilmark{*}, N. Jeremy Kasdin  \altaffilmark{3},  Alexis Carlotti  \altaffilmark{3}, Robert Vanderbei  \altaffilmark{4}}

\altaffiltext{1}{NASA Sagan Fellow, Johns Hopkins University, Department of physics and astronomy,  366 Bloomberg Center 3400 N. Charles Street, Baltimore, MD 21218 USA}

\altaffiltext{2}{Space Telescope Science Institute, 3700 San Marin Drive, Baltimore, MD 21218, USA}
\altaffiltext{3}{Department of Mechanical and Aerospace Engineering, Princeton University, Princeton, NJ, 08544, USA}
\altaffiltext{4} {Department of Operations Research and Financial Engineering, Princeton University, Princeton, NJ, 08544, USA}%

\altaffiltext{*}{Corresponding author: lap@pha.jhu.edu}


\keywords{instrumentation: high angular resolution}

\begin{abstract}

The purpose of this paper is to present the results of a theoretical study pertaining to the feasibility of PIAA units using Deformable Mirrors. We begin by reviewing the general derivation of the design equations driving PIAA. We then show how to solve these equations for square apertures and show the performance of pure PIAA systems in the ray optics regime. We tie these design equations into the study of edge diffraction effects, and provide a general expression for the field after a full propagation through a PIAA coronagraph. Third, we illustrate how a combination of pre and post apodisers yields to a contrast of $10^{-10}$ even in the presence of diffractive effects, for configuration with neither wavefront errors or wavefront control. Finally we present novel PIAA configurations over square apertures which circumvent the constraints on the manufacturing of PIAA optics by inducing the apodisation with two square Deformable Mirrors (DM). Such solutions rely on pupil size smaller than currently envisioned static PIAA solutions and thus require aggressive pre and post-apodizing screens in order to mitigate for diffractive effect between the two mirrors. As a result they are associated to significant loss in performance, throughput in particular.

\end{abstract}

\maketitle 

%
%

\section{Introduction}
\label{sec::intro}

Direct detection and spectral characterization of exo-planets is one of the most exciting challenges of modern astronomy. Over the past decade, the field of high contrast imaging has been extremely active, and a variety of solutions for space based imaging of earth twins have been proposed and tested. Among these concepts, the Phase Induced Amplitude Apodisation coronagraph, first introduced by \cite{2003A&A...404..379G}, is a promising solution since it makes most of the photons collected by the primary mirror of the telescope available for planet detection and characterization. This technique is based on two aspherical mirrors that redistribute the light in the pupil plane of a telescope so that it follows a given amplitude profile, creating a Point Spread Function (PSF) with contrast levels close to $10^{-10}$. Because all the light collected is remapped using these mirrors, such a coronagraph has virtually no throughput loss. As a consequence, the angular resolution is undiminished and is close to the diffraction limit of $1 \lambda/D$, a feature comparable to the performances of a phase mask coronagraph, such as the Optical Vortex Coronagraph (\cite{2010ApJ...709...53M}), but without  any transmissive optics.  Unfortunately, these conclusions are  based on a geometrics optics analysis (\cite{2003ApJ...599..695T},\cite{2006ApJ...639.1129M} ).  \cite{2006ApJ...636..528V} discovered that when a full diffraction analysis is applied to the system, the envisioned designs do not yield the expected level of extinction, but rather, at best, $10^{-5}$ contrast. 

In order to mitigate these diffractive effects, \cite{2005astro.ph.12421P} devised a method using pre -and post-apodisers to retrieve a $10^{-10}$ contrast even while accounting for diffraction. They verified, using a very accurate numerical integrator, that apodised PIAA coronagraphs could be designed to have performances close to the ones predicted by ray optics.  In the absence of apodizing screens, another avenue to mitigate for propagation induced artifacts is to treat them as wavefront errors and compensate for them with one or two Deformable Mirrors (\cite{2007Natur.446..771T}, \cite{Pueyo:09},\cite{2010PASP..122...71G}). The first solution, apodizing screens, will reduce the throughput of the coronagraph, and thus the efficiency of the instrument, but preserve the spectral bandwidth. The second alternative, wavefront control, will preserve throughput but lead to a reduced spectral bandwidth since the diffractive oscillations are highly chromatic and cannot be fully compensated  by deformable mirrors. In this communication we treat only the  former design approach, where diffractive  effects are mitigated by pre- and post-apodizers.

In this paper we extend the theoretical framework of \citet{2006ApJ...636..528V} and \citet{2005astro.ph.12421P} by developing an analytical formalism for  the diffractive effects on arbitrary geometries.
These results are not only useful for a qualitative understanding of pupil mapping coronagraphs, but are also critical for the design of such solutions, \cite{balasubramanian:77314U}, for numerical studies pertaining to their sensitivities to aberrations, \cite{Pueyo:11}, and for understanding how to integrate with wavefront control systems, \cite{krist:77314N}. We apply these findings to pursue alternate designs of PIAA coronagraphs, and focus on the particularly interesting case of square apertures. Indeed, if  the apodisation function is written as the product of two separable, one-dimensional functions, \cite{2001ApJ...548L.201N}, very deep contrast can be achieved on the four diagonal directions of the Point Spread Function (PSF) with only mildly deformed PIAA mirrors. This configuration implies that if the telescope aperture is circular, the square sub-aperture needs to be inscribed in the telescope pupil, leading therefore to a throughput loss of a factor of $2/\pi $.  Since any exo-planet dedicated coronagraph will have to be equipped with two sequential DMs to reduce the stellar halo due to imperfection in the optics at mid-spatial frequencies (\cite{2006ApOpt..45.5143S,Pueyo:09}), and since current state of the art piezo-electric or MEMS (Microelectromechanical Systems) devices all have a square formats (\cite{2007Natur.446..771T}, \cite{Giveon:07}), it is sensible to study square apertures inscribed in a circular telescope aperture. Alternatively, one could envision a design for which the circular pupil is inscribed in a square deformable mirror. This would increase the net throughput of the system, and the Inner Working Angle (IWA) would become constant for all azimuthal angles of the PSF. However, the Outer Working Angle (OWA) would not take advantage of all the actuators on the Deformable Mirror in the diagonal direction and the PIAA deformations would become larger.  Adressing such a paradigm is beyond the scope of this paper, and we focus  in this communication on the case where the square pupil is  inscribed in the circular aperture, the case that is widely implemented in current high contrast testbeds (\cite{2007Natur.446..771T}, \cite{Pueyo:09},\cite{2010PASP..122...71G}).

We begin by reviewing the general derivation of the defining PIAA equations and show how to solve these equations in the simplified case of square separable apertures.  We present the instrument response in that case, and study the performance of this pure PIAA systems under the ray optics approximation.  In a second section we revisit the analysis considering  edge diffraction effects and 
provide a general expression for the field after a full propagation through any arbitrary PIAA coronagraph. In the remainder of the paper we apply these general results to the design and trade-offs associated with PIAA coronagraphs over square separable apertures. We present numerical results that use our new diffractive approximation and illustrate how a combination of pre- and post-apodisers restores contrast below $10^{-10}$ even in the presence of diffractive effects. Finally, using the result of this theoretical study, we present novel PIAA designs over square apertures which can potentially relax the constraints on the manufacturing of PIAA optics or completely circumvent manufacturing by inducing the apodisation with two square Deformable Mirrors (DM).

\section{Design of Pupil remapping systems}

\subsection{General equations}

The Phase Induced Amplitude Apodization (PIAA) coronagraph was fist introduced by \citet{2003A&A...404..379G} as an optical apparatus composed of two reflective surfaces that remapped the distribution of the light in the pupil of  a telescope into an apodisation function which yields a high contrast PSF. \citet{2003A&A...404..379G} showed that the mirror surfaces could be derived as  the solution of a single differential equation and \cite{2003ApJ...599..695T} subsequently developed this result showing that the relationship between prescribed apodisation and mirrors surfaces could also be written as a set of two coupled partial differential equations. Here we rederive these ray-optics based design equations for arbitrary geometries and illustrate how to solve them using square apertures.

\begin{figure}
\includegraphics[width=6in,angle=-90]{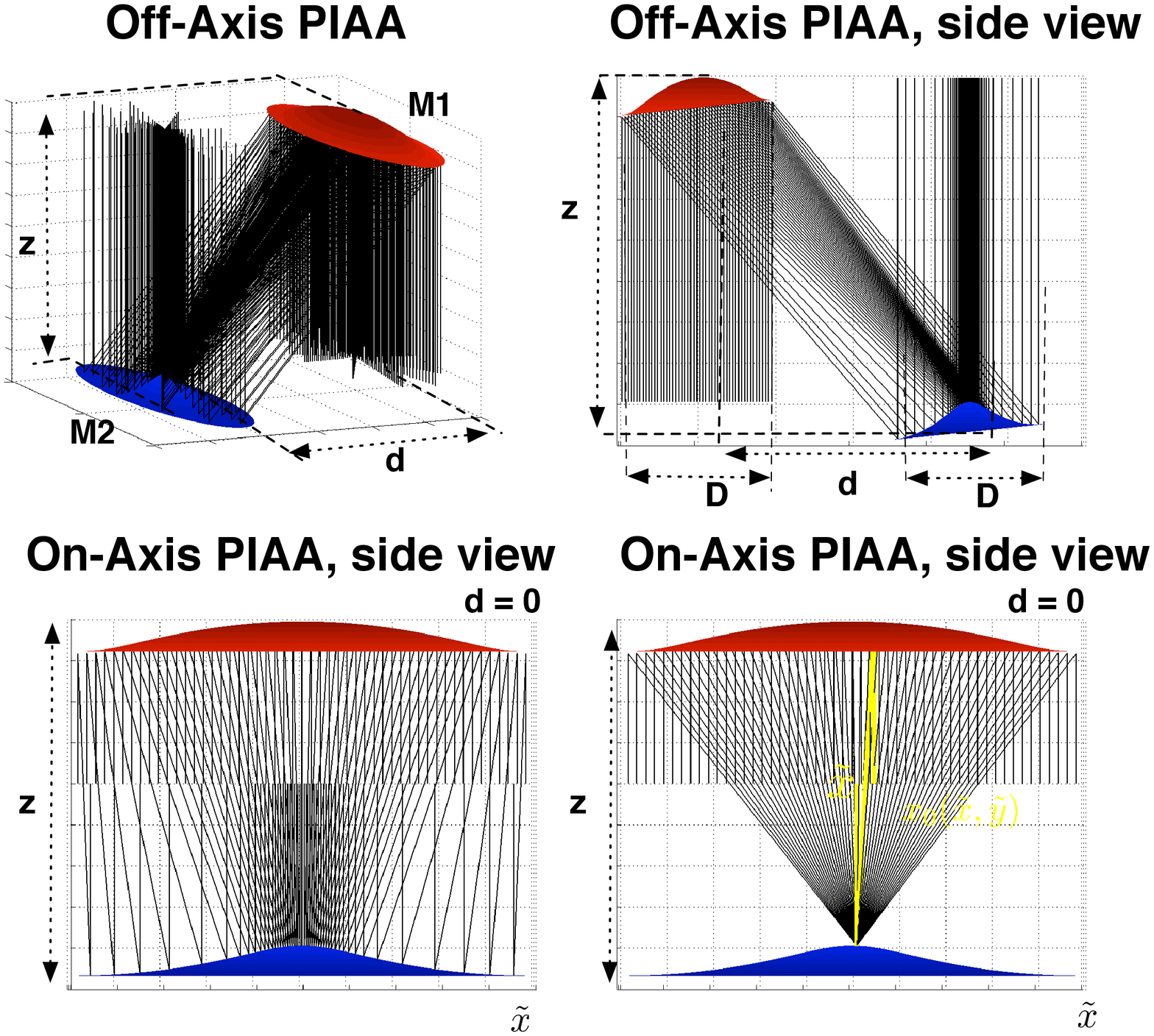}
\caption[Geometry and notation]{Geometry and notation. Top Left: Three dimensional representation of a pupil-to-pupil off-axis PIAA system. Top Right: Side view of the geometrical remapping in a pupil-to-pupil off-axis PIAA system. Bottom Left: Side view of the geometrical remapping
in a pupil-to-pupil on-axis PIAA unit: \textbf{This is the configuration that is studied in this communication}. Bottom right: Side view of all the rays contributing to the diffractive  field at a point of coordinates $(\tilde{x},\tilde{y})$ at M2. The ray corresponding to the geometrical remapping, which has coordinates $(x_i(\tilde{x},\tilde{y}), y_i(\tilde{x},\tilde{y}))$ in the input plane, is highlighted.}
\label{FigSetupPIAA}
\end{figure}

Consider the arrangement and coordinate systems shown in Fig.~\ref{FigSetupPIAA}. Note that in this paper we only consider the on-axis pupil-to-pupil configurations leaving aside the off-axis focal-to-focal case, whose surfaces cannot be derived by direct integration of the systems of partial differential equations. For now we do not assume anything about the geometry of  the two remapping mirrors: $M1$ and $M2$ are two domains of $\mathbb{R}^2$ of arbitrary shapes. We introduce the independent variables $(x,y)$ that span the domain of $M1$ and $(\tilde{x},\tilde{y})$ that span the domain of M2.  The height of M1 is thus given by the function $h(x,y)$ and the height of M2 is given by $\tilde{h}(\tilde{x},\tilde{y})$.

Given the distance between the two mirrors, $Z$, the optical path length, $Q$, between any two points $(x,y)$ and $(\tilde{x},\tilde{y})$ is  given by
%
%
\begin{eqnarray}
Q(x,y,\tilde{x},\tilde{y}) &=& h(x,y)-\frac{Z}{2}+S(x,y,\tilde{x},\tilde{y})+\frac{Z}{2}-\tilde{h}(\tilde{x},\tilde{y}) \nonumber \\
 &=&S(x,y,\tilde{x},\tilde{y})+h(x,y)-\tilde{h}(\tilde{x},\tilde{y})
\label{DefOPL}
\end{eqnarray}
where $S$ is the free space propagation within the apparatus
\begin{equation}
S(x,y,\tilde{x},\tilde{y})=\sqrt{(x-\tilde{x})^{2}+(y-\tilde{y})^{2}+(h(x,y)-\tilde{h}(\tilde{x},\tilde{y}))^{2}} .
\label{DefS}
\end{equation}

Following a two-dimensional generalization of the approach in \cite{2005ApJ...626.1079V}, the pupil mapping solution for the mirror heights is a one-to-one mapping between points in the domain of M1 to points in the domain of M2.
For instance, if we consider the coordinates $(x,y)$ on M1 as independent variables, then the pupil mapping solution maps rays to specific points $(\tilde{x}_o,\tilde{y}_o)$ via two functions
\begin{eqnarray}
\tilde{x}_o &=& f_1(x,y) \label{eq:f1} \\
\tilde{y}_o &=& f_2(x,y). \label{eq:f2}
\end{eqnarray}
Alternatively, since the pupil mapping solution must be reversible, we can consider the points $(\tilde{x},\tilde{y})$ as independent variables and find the inverse mapping,
\begin{eqnarray}
x_i &=& g_1(\tilde{x},\tilde{y}) \label{eq:g1} \\
y_i &=& g_2(\tilde{x},\tilde{y}) \label{eq:g2}
\end{eqnarray}
where $g_1$ and $g_2$ are the inverse functions of $f_1$ and $f_2$.  Typically we design the pupil mapping system by selecting the inverse functions $g_1(\tilde{x},\tilde{y})$ and $g_2(\tilde{x},\tilde{y})$ since our goal is a specific amplitude profile at M2.

%
We design a pupil remapping unit via ray optics by choosing the heights of each mirror, $h(x,y)$ and $\tilde{h}(\tilde{x},\tilde{y})$, so that the output amplitude profile follows the desired apodisation $A(\tilde{x},\tilde{y})$.  This requires that the following three constraints be satisfied:
\begin{enumerate}
\item  \label{energy_conservation} The intensity profile after $M2$ matches the square of the desired apodisation $A(\tilde{x},\tilde{y})^2$. The relationship between this profile and the mirror surfaces has to satisfy local energy  conservation between $M1$ and $M2$.

\item \label{Fermat} For each ray between $(x_i,y_i)$ and $(\tilde{x}_o,\tilde{y}_o)$,  Fermat's principle must be satisfied. 

\item \label{constant_phase} When the apparatus is illuminated by uniform on-axis light, the outgoing wavefront is flat. This means that, in the absence of diffractive effects, the phase at $M2$ is constant when the wavefront hitting $M1$ is flat.
\end{enumerate} 
In the remainder of this section we show how, for any arbitrary geometry, these constraints translate into formal conditions on the heights of each mirror under a ray optics approximation. The set of equations obtained defines the design of a PIAA unit.\\

\subsubsection{Energy remapping}
\begin{figure}[b]
\begin{tabular}{cc}
\includegraphics[width=.4\columnwidth]{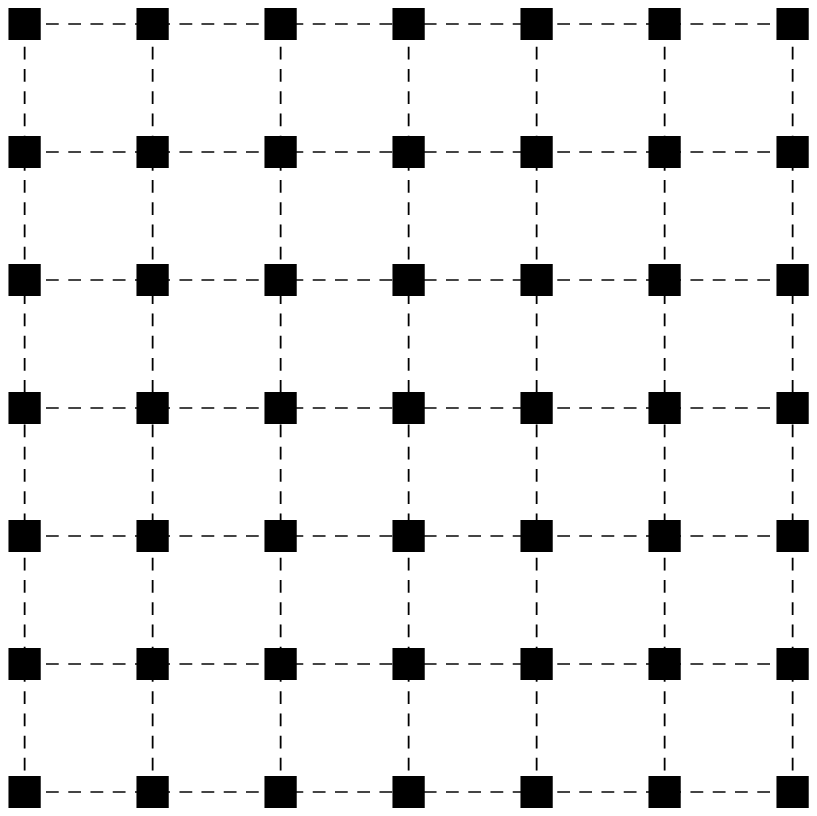} & \includegraphics[width=.4\columnwidth]{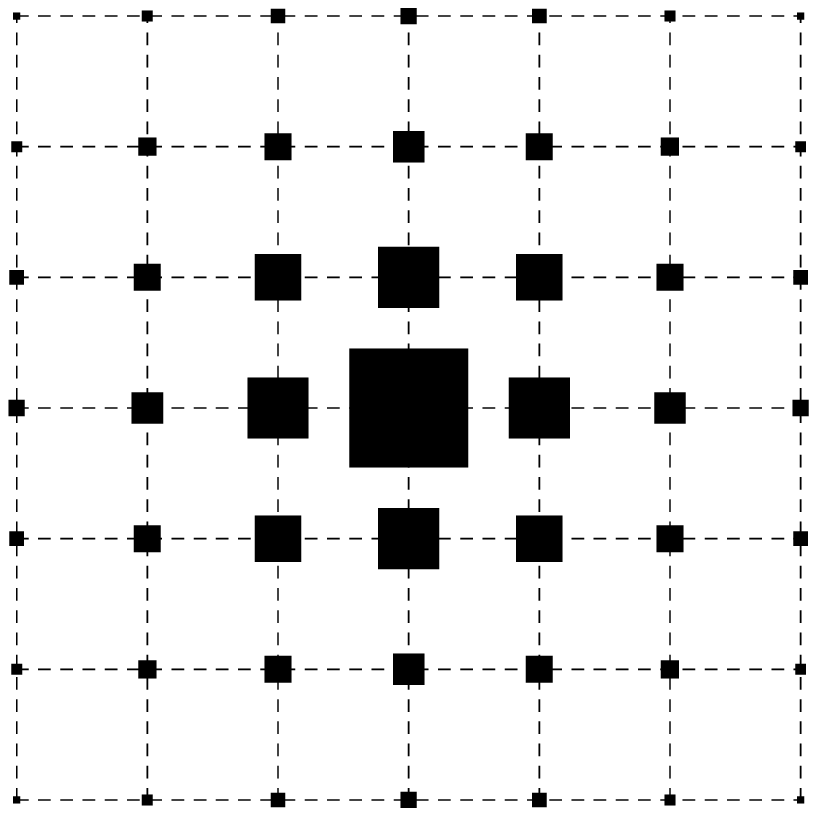}
\end{tabular}
\caption{Illustration of the pupil mapping effect on the light intensity distribution. Left: the energy distribution is uniform on $M1$. Right: the energy distribution on $M2$. The area of each square is then proportional to the intensity of the apodisation function $A(\tilde{x},\tilde{y})^{2}$.
\label{FigEnergyRemapp}}
\end{figure}

Constraint~\ref{energy_conservation} is simply a statement of energy conservation.
 This translates directly into an equation for the remapping functions $g_1$ and $g_2$ (or, equivalently, $f_1$ and $f_2$).  
Fig. \ref{FigEnergyRemapp} illustrates that the energy in a small square surface element of $M1$ must equal the energy in its remapped image at $M2$. Because of the aspherical nature of the PIAA mirrors the area of the each element varies as a function of  location at $M2$ and thus, in order to conserve energy, the light intensity in each square changes accordingly.  More formally, this implies that the energy contained in a differential area $dx dy$ at $M1$ must remain constant when traveling through the set of aspherical PIAA optics.  Considering $(\tilde{x},\tilde{y}$) as independent variables, we can view the inverse mapping  as a coordinate transformation, given by Eqs.~\ref{eq:g1} and \ref{eq:g2}, allowing us to use the differential area formula,
\begin{equation}
dx dy =  \det( J\left(\tilde{x},\tilde{y}) \right) d \tilde{x} d\tilde{y}\\
\label{eq:Jacobian}
\end{equation}
%
where $J(\tilde{x},\tilde{y})$ is the Jacobian matrix of the change of coordinates  $(\tilde{x},\tilde{y})$ to $(x_i,y_i)$.  Since the remapped rays correspond to an amplitude profile $A(\tilde{x},\tilde{y})$, the energy in a bundle of rays at $M2$ is given by $A(\tilde{x},\tilde{y})^2$.  Consequently, the energy remapping constraint in Eq.~\ref{eq:Jacobian} can be written as
\begin{equation}
\frac{\partial g_1}{\partial \tilde{x}} \frac{\partial g_{2}}{\partial \tilde{y}} - \frac{\partial g_{1}}{\partial \tilde{y}}\frac{\partial g_{2}}{\partial \tilde{x}} = A(\tilde{x},\tilde{y})^2
\label{EqRemap}
\end{equation}
where, again, $g_1(\tilde{x},\tilde{y})$ and $g_2(\tilde{x},\tilde{y})$ are the remapping functions in Eqs.~\ref{eq:g1} and \ref{eq:g2}.

The desired apodisation thus does not directly lead to the mirror surface heights but only to the location of the incident rays. This implies, as we will study in the next section, that the output field strongly depends on how well the ray optics approximation holds. Eq.~\ref{EqRemap} is a two-dimensional version of the well known non-linear partial differential equation known as the Monge-Ampere equation, \cite{MongeVieux}. While there is no known general solution to this equation for  $g_1$ and $g_2$, we show that, by using a square aperture we can find solutions if we restrict the apodisation to be the tensor product of two one-dimensional functions. 

\subsubsection{Mirror Heights}
To find the heights of M1 and M2 we follow the approach in \cite{2005ApJ...626.1079V} and satisfy Fermat's principle (constraint \ref{Fermat}) by minimizing the optical path length of each ray from $(x_i,y_i)$ to $(\tilde{x}_o,\tilde{y}_o)$.  
After some algebraic manipulations this provides the equation governing the height of $M1$ at each $(x,y)$ across the mirror,
%
\begin{eqnarray}
\frac{\partial h}{\partial x} (x,y) &=& \frac{f_1(x,y)-x}{S(x,y,f_1,f_2)+\big (h(x,y)-\tilde{h}(f_1,f_2)\big )} \label{Eqhx} \\
\frac{\partial h}{\partial y} (x,y)) &=& \frac{f_2(x,y)-y}{S(x,y,f_1,f_2)+\big (h(x,y)-\tilde{h}(f_1,f_2)\big )} .
\label{Eqhy}
\end{eqnarray}
%
%
%
%

For the height of M2 we use constraint \ref{constant_phase}, which states that the phase of the wavefront be constant across the exit pupil.  It was shown in \cite{2005ApJ...626.1079V} that this translates into the condition that each ray between the pupil remapped points $(x_i,y_i)$ and $(\tilde{x}_o,\tilde{y}_o)$ must have constant optical path length.  From Eq.~\ref{DefOPL} this means that
\begin{eqnarray}
Q_0 &=& S(g_1,g_2,\tilde{x},\tilde{y}) +\left (h(g_1,g_2)-\tilde{h}(\tilde{x},\tilde{y})\right ) \nonumber  \\
 &=& S(x,y,f_1,f_2) +\left (h(x,y)-\tilde{h}(f_1,f_2)\right ) \nonumber \\
&=& 2Z . \label{eq:constant_Q}
\end{eqnarray}

Taking the partial derivative of the optical path length with respect to $(\tilde{x},\tilde{y})$ and setting it equal to zero leads to expressions for the height of M2, 
%
\begin{eqnarray}
\frac{\partial \tilde{h}}{\partial \tilde{x}}(\tilde{x},\tilde{y})  &=& \frac{\tilde{x}-g_1(\tilde{x},\tilde{y})}{S(g_1,g_2,\tilde{x},\tilde{y})+\big (h(g_1,g_2)-\tilde{h}(\tilde{x},\tilde{y})\big )} \label{Eqtildehx} \\
\frac{\partial \tilde{h}}{\partial \tilde{y}}(\tilde{x},\tilde{y}) &=& \frac{\tilde{y}-g_2(\tilde{x},\tilde{y})}{S(g_1,g_2,\tilde{x},\tilde{y})+\big (h(g_1,g_2)-\tilde{h}(\tilde{x},\tilde{y})\big )} . \label{Eqtildehy}
\end{eqnarray}
%
%

%
The resulting set of partial differential equations defining a pupil mapping system thus reduces to:
\begin{eqnarray}
\frac{\partial g_1}{\partial \tilde{x}} \frac{\partial g_{2}}{\partial \tilde{y}} - \frac{\partial g_{1}}{\partial \tilde{y}}\frac{\partial g_{2}}{\partial \tilde{x}} &=& A(\tilde{x},\tilde{y})^2 \label{EqRemapBis}\\
\frac{\partial h}{\partial x}& =& \frac{f_1(x,y)-x}{2Z} \label{EqhxS} \\
\frac{\partial h}{\partial y}& =& \frac{ f_2(x,y)-y}{2Z} \label{EqhyS}\\
\frac{\partial \tilde{h}}{\partial \tilde{x}}& =& \frac{\tilde{x}-g_1(\tilde{x},\tilde{y})}{2Z} \label{EqtildehxS} \\
\frac{\partial \tilde{h}}{\partial \tilde{y}}& =& \frac{ \tilde{y}-g_2(\tilde{x},\tilde{y})}{2Z} \label{EqtildehyS} .
\end{eqnarray}
where $f_1$ and $f_2$ are the inverse functions associated with $g_1$ and $g_2$.
Variations of these equations were first derived by \cite{2005ApJ...626.1079V} for circular geometries using ray optics. They are the governing relationships for the design of a PIAA unit.

While it is possible to design pupil remappers with two mirrors of different areas, we will restrict our study here to area preserving systems, meaning the set  $\mathcal{S}$ of $(x,y)$ over the entrance pupil is the same as the set of $(\tilde{x},\tilde{y})$ over the exit pupils.

Note that, as shown in \cite{2005ApJ...626.1079V}, it is sufficient to solve Eqs.~\ref{EqRemapBis}, \ref{Eqtildehx}, and \ref{Eqtildehy} for $g_1(\tilde{x},\tilde{y})$, $g_2(\tilde{x},\tilde{y})$, and $\tilde{h}(\tilde{x},\tilde{y})$, since, by conservation of optical path length,
\begin{equation}
h(x,y) = \tilde{h}(f_1(x,y),f_2(x,y)) - \frac{Z}{2} + \frac{(x - f_1(x,y))^2 + (y-f_2(x,y))^2}{2 Z}
\end{equation}
where $f_1$ and $f_2$ are the inverse of $g_1$ and $g_2$.  This conservation of optical path reduces the problem to solving only one second order partial differential equation (by solving Eq.~\ref{Eqtildehx} for $g_1$, Eq.~\ref{Eqtildehy} for $g_2$ and substituting into Eq.~\ref{EqRemapBis}). This finishes our derivation of the design equations of PIAA systems using ray optics. We now focus on applying them to the case of separable apodization functions on square apertures. 

\subsection{Mirror shapes for a square geometry}
As we mentioned above, there is no general solution to the Monge-Ampere equation and thus no closed form pupil mapping design for arbitrary pupil geometries (\cite{MongeVieux} \cite{Glimm02}). \cite{2005ApJ...626.1079V}  restricted themselves to a circular geometry, which matches a typical telescope entrance pupil.  The resulting azimuthal symmetry reduces the pupil mapping equations to one-dimensional partial differential equation's in radius, leading to relatively simple solutions for the mirror profiles.  We consider here the case of a square aperture inscribed in the circular telescope pupil.

To make the design tractable on square pupils we further reduce the problem by forcing a separability condition on the prescribed apodisation,
\begin{equation}
A(\tilde{x},\tilde{y})^2 = A_X(\tilde{x})^2 A_Y(\tilde{y})^2.
\end{equation} 
Under this constraint, there are multiple solutions of Eqs. \ref{Eqtildehx} and \ref{Eqtildehy}. We  thus choose to search for $g_1$, $g_2$ and $\tilde{h}$ such that:
\begin{eqnarray}
g_1(\tilde{x},\tilde{y}) &=& g_1(\tilde{x} \label{EqXsep})\\
g_2(\tilde{x},\tilde{y}) &=& g_2(\tilde{y} \label{EqYSep})\\
\tilde{h}(\tilde{x},\tilde{y}) &=& \tilde{h}_X(\tilde{x})+ \tilde{h}_Y(\tilde{y}) \label{EqHsep}
\end{eqnarray}
Note that Eq.~\ref{EqHsep} can be deduced directly from Eqs.~\ref{EqXsep} and \ref{EqYSep}. 
These constraints reduce Eq.~\ref{EqRemapBis} to the separable form,
\begin{equation}
\frac{\partial g_1}{\partial \tilde{x}} \frac{\partial g_2}{\partial \tilde{y}}  =  A_X(\tilde{x})^2A_Y(\tilde{y})^2. \label{EqRemapSep}
\end{equation}
Considering a pupil of width $a$, we use the facts that both area and apertures size are conserved, namely $\int A_X(\tilde{x})^2 d \tilde{x} = a$ , or $\int  d x_{i}(\tilde{x}) = a$, and we find, via separation of variables, 
\begin{eqnarray}
g_1(\tilde{x}) &=& \int_{-a/2}^{\tilde{x}} A_X(\sigma)^2 d \sigma -\frac{a}{2}\\
g_2(\tilde{y}) &=& \int_{-a/2}^{\tilde{y}} A_Y(\sigma)^2 d \sigma -\frac{a}{2}\\
\tilde{h}(\tilde{x},\tilde{y}) &=& \frac{1}{Z} \int_{-a/2}^{\tilde{x}} \left( \int_{-a/2}^{\nu} A_X(\sigma)^2 d \sigma - \nu  \right) d \nu + \frac{1}{Z} \int_{-a/2}^{\tilde{y}} \left( \int_{-a/2}^{\tau} A_Y(\sigma)^2 d \sigma - \tau  \right) d \tau 
\end{eqnarray}
%

\begin{figure}[htbf]
\includegraphics[width=2.4in,angle=-90]{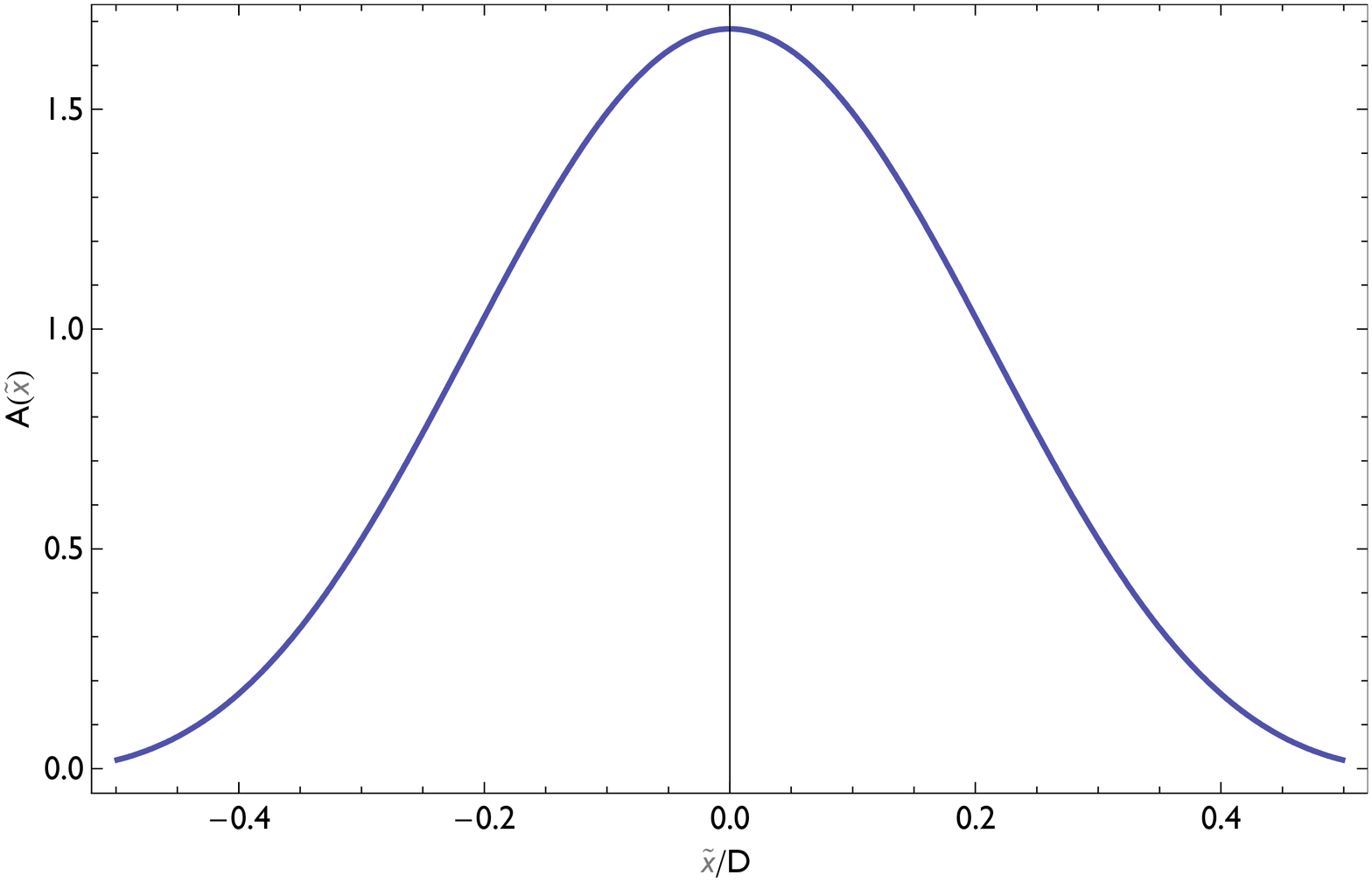}
\includegraphics[width=2.4in,angle=-90]{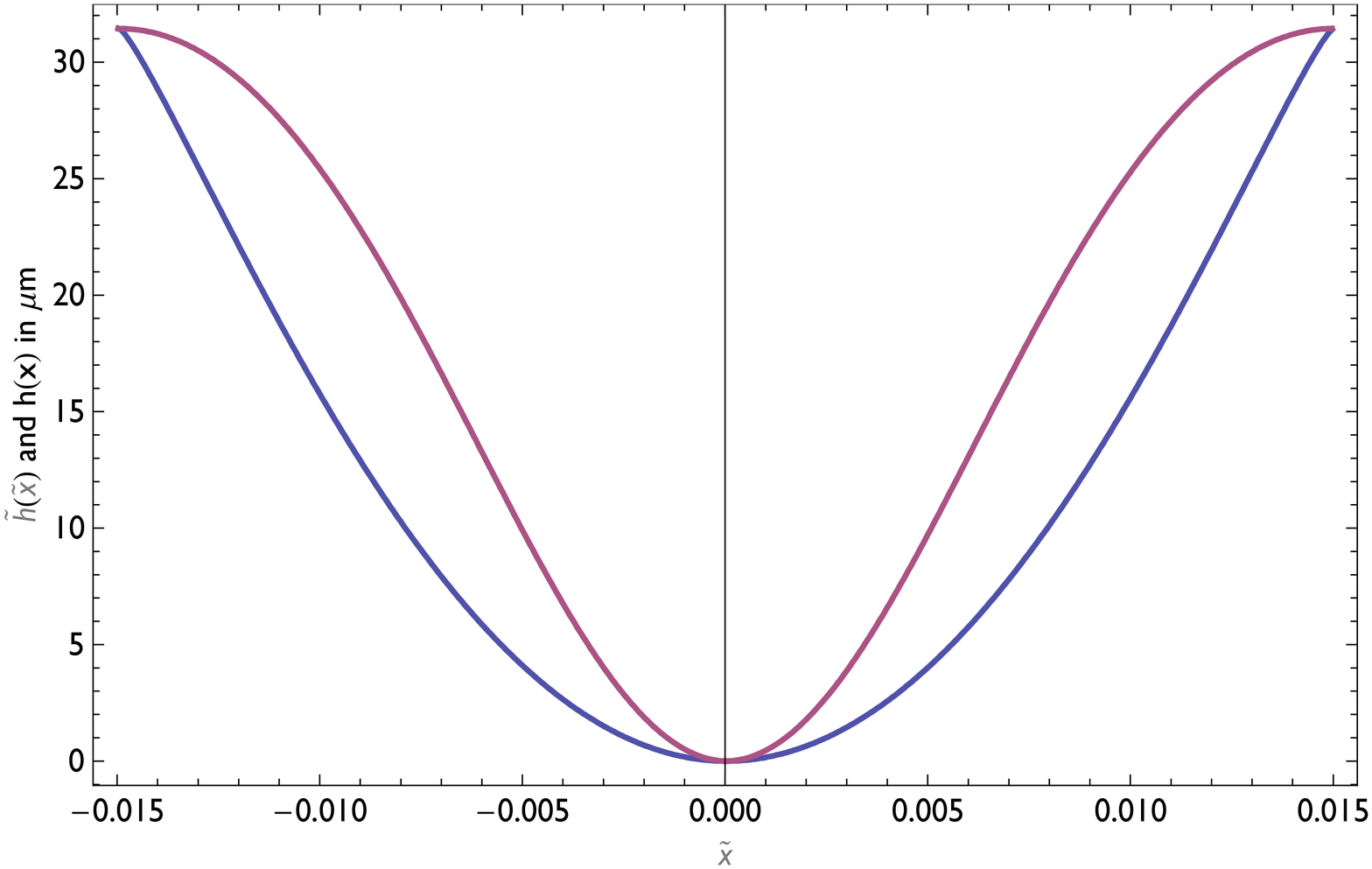}
\caption[Mirror shapes in the case of square apertures.]{Apodisation in square apertures. Left: One dimensional apodisation. Right: One dimensional shape of $M1$ and $M2$. Aperture size $D= 3$ cm, mirrors distance $z=1$ m}
\label{FigShapesPIAACart}
\end{figure}

\begin{figure}[htbf]
\begin{center}
\includegraphics[width=5in,angle=-90]{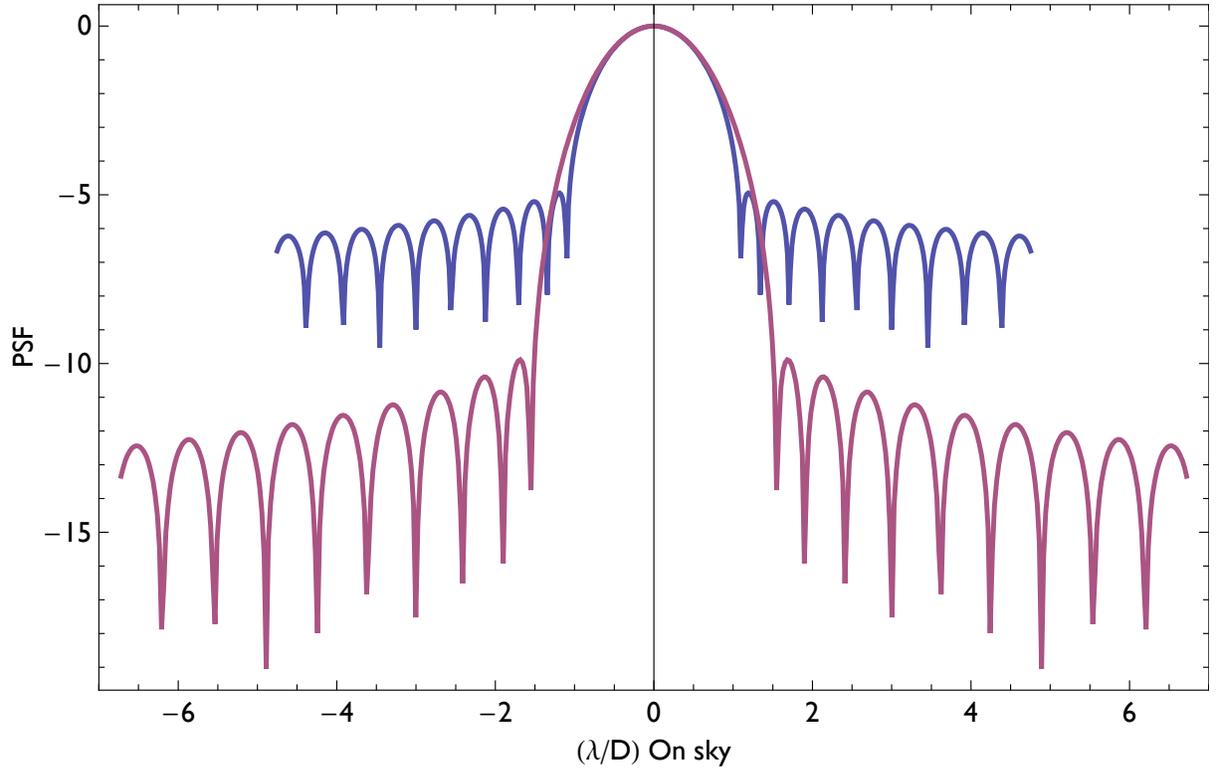}
\caption[PSF in the case of square apertures.]{One dimensional profiles along the horizontal axis and the diagonal of the square aperture PSF. The x axis is in units of $\lambda /D _{On Sky}$. This PSF was computed assuming that ray optics hold perfectly.}
\label{PSF1DNoApod}
\end{center}
\end{figure}

We can apply this result to design a PIAA that reproduces an optimal high contrast apodisation over square apertures. The idea of using square apertures and taking advantage of the higher contrast on the diagonal of the PSF was first introduced by \cite{2001ApJ...548L.201N}, who used sonine and cosine apodisations. This idea was re-visited by \cite{2001A&A...379..697A} who introduced an interferometric induced apodisation. They also pointed towards using optimal functions for square apertures, called prolate functions (\cite{2002A&A...389..334A}). Here we use a combination of a horizontal and vertical cross prolate apodizer. Each apodizing screen yields a $10^{-5 }$ contrast, resulting in a constrast on the diagonal of the PSF of $10^{-10}$. Note that this is also the underlying idea behind the design of checkerboard shaped pupils (\cite{2004ApJ...615..555V}). Using prolate apodizations in each orthogonal direction, we thus find two mirror shapes, Fig.~\ref{FigShapesPIAACart}, which lead to a PSF that achieves high contrast in a region around each diagonal, Fig.~\ref{PSF1DNoApod}. Note that the deformations along the axes of the square apertures are smaller magnitude than in the azimuthally symmetric case, which seems natural since separable square apertures do not present a full $360$ degrees search space.\\

%
\subsection{Performance}
\begin{figure}[htbf]
\begin{center}
\includegraphics[width=2.8in,angle=-90]{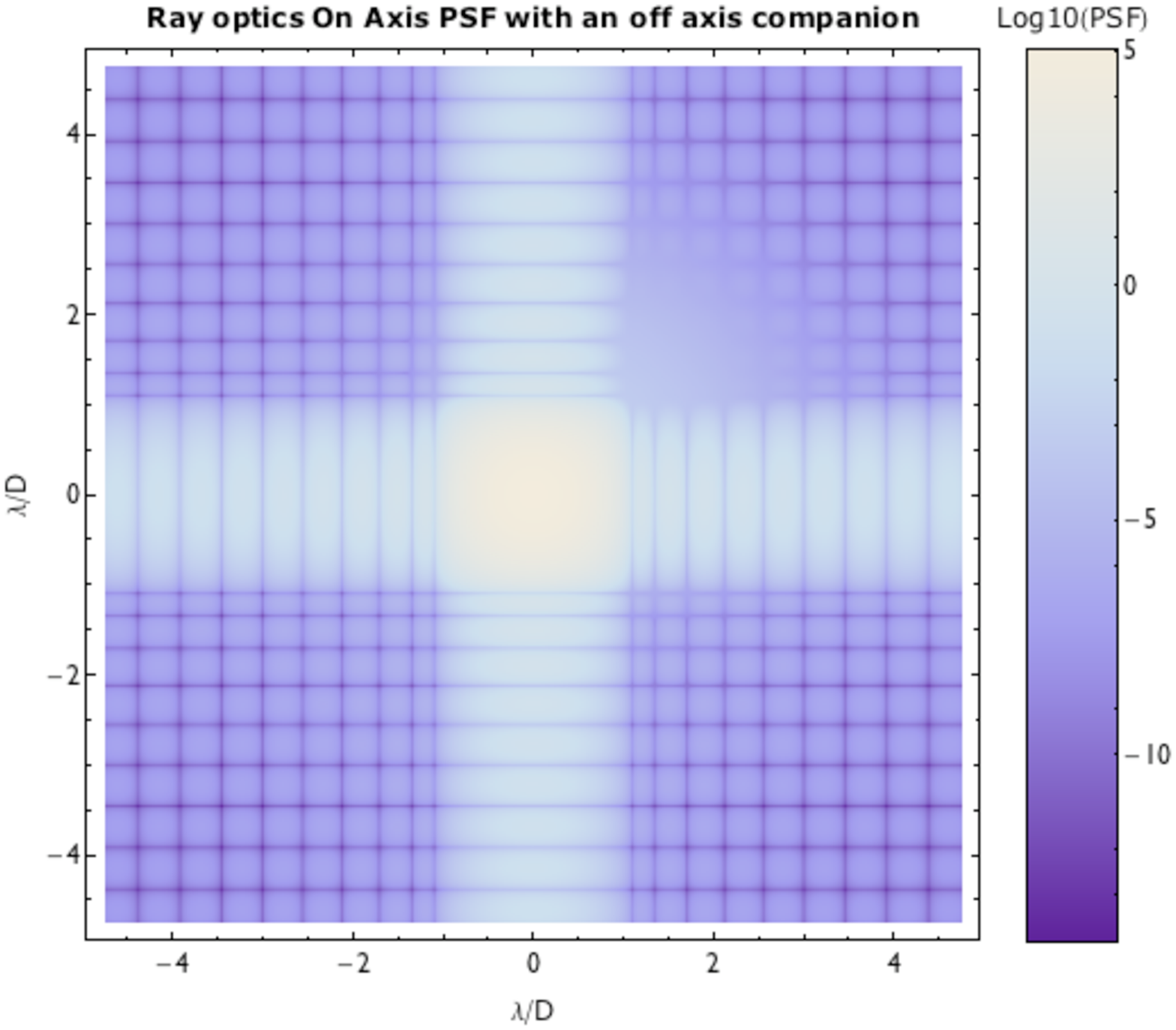}
\includegraphics[width=2.8in,angle=-90]{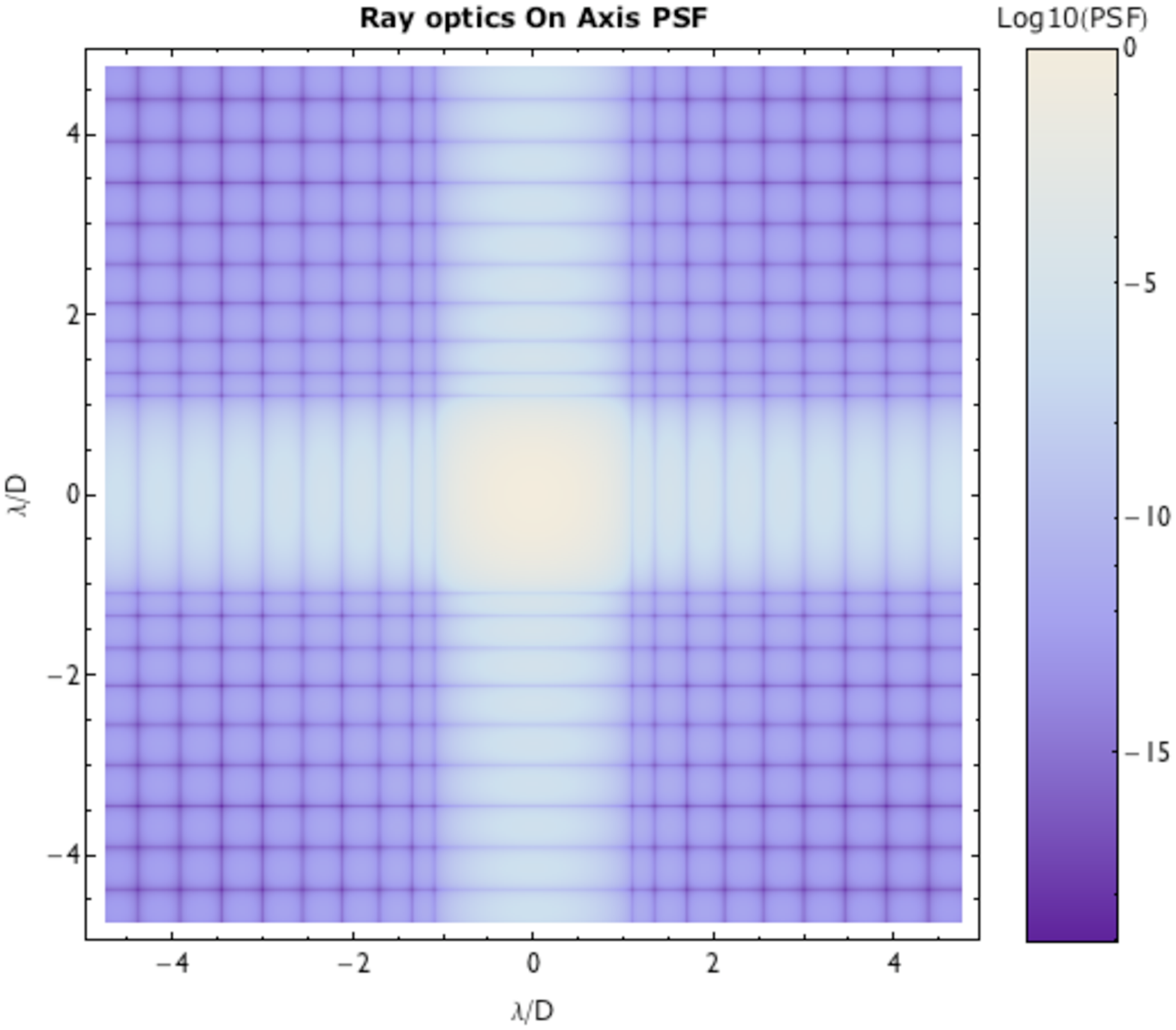}
\caption[PSF in the case of square apertures.]{Case of square apertures. Top: Two dimensional PSF. This PSF was computed assuming that ray optics hold perfectly. Bottom: A companion of brightness $10^{-9}$ was added to the top right corner at $2 \lambda / D$}
\label{PSFNoApod}
\end{center}
\end{figure}

Except for the throughput loss associated with  inscribing the pupil in a square, a square pupil mapper has all of the advantages of classical PIAA coronagraphs.  When designed using only ray optics it achieves the needed contrast with nearly $100$ percent throughput.  Moreover, because of the unique combination of the close inner working angle of square designs and the magnification effects of a pupil mapper system, described in Appendix A, it has a very small inner working angle, as close as $1.5 \lambda/D$.  This magnification is due to the dependence of optical path length on the angle of incidence.  
Our pupil mapping design achieves a contrast of $10^{-10}$ with an inner working angle (IWA) of $1.8 \lambda/D$.  However, like other square pupil apodisations, this is not achieved everywhere but only in the region around the diagonals covering roughly $63$ percent of the image plane. This is illustrated on Fig.~\ref{PSFNoApod} where on-axis and off-axis two dimensional PSFs are shown.

Unfortunately, like all pupil mapping approaches, this ray optics performance is not achievable in practice. Contrast is limited by light diffracted off the edges of $M1$. Under narrow band illumination the $10^{-10}$ contrast can be retrieved using a Deformable Mirror based wavefront controller that treats the electrical field oscillations due to the propagated edges of $M1$ as any other chromatic wavefront error. However, because the magnitude of such oscillations  varies rapidly and periodically with wavelength, it is fundamentally difficult for wavefront control technologies to correct them down to high contrast levels, eg $10^{-10}$, over a broad spectral bandwidth, $\Delta \lambda / \lambda > 0.05$. Here we choose to explore a solution, based on apodizing screens,  which yields high contrast over a broadband but at a cost in throughput. Such designs are intrinsically achromatic and transfers the problem of bandwidth to the control of wavelength dependent optical aberration, \cite{Pueyo:09,2007ApJ...666..609P}. In the next section we present an analytical treatment of the edge-diffraction properties of PIAA designs.

\section{Diffraction Analysis}

The previous section derived the  pupil remapping equations using only ray optics.  As we noted earlier, it has already been shown for circular geometries (\cite{2005ApJ...626.1079V}) that when a careful diffraction analysis is carried out, the contrast yielded by a pupil mapping system is degraded by many orders of magnitude.  In this section we present a new approximation to the diffraction integral on arbitrary geometries and show how it leads to the pupil mapping apodization in the ray optics limit. This approximation is geometry independent and has multiple applications beyond the scope of this communication. In particular it has been applied to design robust and efficient numerical propagators of arbitrary electrical fields through PIAA coronagraphs, \cite{krist:77314N}, a tool that is critical for the design and diagnostic of experimental validations.

%
\subsection{The Main Theorem}
The exact electric field at the exit pupil of a PIAA unit can be found using the Huygens integral over all the possible optical paths from $(x,y)$ at M1, to a point $(\tilde{x},\tilde{y})$ in the output plane, after M2,
\begin{equation}
E_{out}(\tilde{x},\tilde{y})=\frac{E_0}{2 i \lambda Z} \int \int e^{\frac{2i \pi}{\lambda}Q(x,y,\tilde{x},\tilde{y})} dx dy  \label{huygens}
\end{equation}
where $E_0$ is the constant entrance electric field, $Q(x,y,\tilde{x},\tilde{y})$ is the optical path length between any two points in the entrance and exit pupils given by Eq.~\ref{DefOPL}. We have used the paraxial approximation to replace the optical path length in the denominator with $2Z$.  Unfortunately, this integral is intractable for analysis and computation.  The classical Fresnel approximation expands the optical path length in the exponential about its value along the axis of the optical system ($x=0$ and $y=0$).  As was shown in \cite{2005ApJ...626.1079V}, this approximation is too crude for a pupil mapping analysis.  Instead, we expand the optical path length about the constant $Q_0$ along the ray between the pupil remapped coordinates,  $(\tilde{x}, \tilde{y})$ and $x_i=g_1(\tilde{x},\tilde{y})$, $y_i=g_2(\tilde{x},\tilde{y})$ in Eq.~\ref{eq:constant_Q},
\begin{eqnarray}
Q(x,y,\tilde{x},\tilde{y}) & \cong & Q_0  +  \frac{1}{2}\frac{\partial^{2} Q}{\partial x^{2}}\big |_{x=x_i,y= y_i} (x-x_i)^{2} \nonumber \\
&+& \frac{\partial^{2} Q}{\partial x \partial y}\big |_{x=x_i, y=_i}(x-x_i)(y-y_i) \nonumber \\
&+& \frac{1}{2}\frac{\partial^{2} Q}{\partial y^{2}}\big |_{x=x_i, y=y_i}(y-y_i)^{2}.
\end{eqnarray}
Note that the first order derivatives vanished because the functions $g_1$ and $g_2$ where chosen to minimize the optical path length.  This means we can write the Huygens integral in Eq.~\ref{huygens} as
%
\begin{equation}
E_{out}(\tilde{x},\tilde{y})= \frac{E_0e^{\frac{2 i \pi}{\lambda}Q_0}}{2 i \lambda Z} \int \int e^{\frac{2 i \pi}{\lambda} G_{2}(x,y,\tilde{x},\tilde{y})} dx dy
\label{EqTaylorHuygensShort}
\end{equation}
where $G_2$ only contains terms of the second order, $(x-x_i)^2, (x-x_i)(y-y_i), (y-y_i)^2$. We assume here that the higher order terms are negligible, using the same arguments presented in \cite{Goodman}. 
The field after M2 is thus mainly described by an integral over the quadratic terms of the Taylor expansion of the optical path length across the apparatus.  This brings us to our main theorem:
%
%
\begin{theorem}
\label{MainPropThorem}
Consider an arbitrary pupil remapping system designed such that the geometric PIAA equations are satisfied. Then, as long as the curvature on the mirrors is smooth enough, the propagated field after $M2$ can be approximated as:
\begin{equation}
E_{out}(\tilde{x},\tilde{y}) \cong \frac{e^{\frac{2 i \pi}{\lambda} Q_0}}{2 i \lambda Z} \int \int e^{\frac{ i \pi }{\lambda S_0} \big( \dpart{f_1}{x}\big |_{x_i,y_i}(x-x_i)^{2} + 2 \dpart{f_2}{x}\big |_{x_i,y_i}(x-x_i)(y-y_i) +  \dpart{f_2}{y}\big |_{x_i,y_i} (y-y_i)^{2} \big) } dx dy
\label{EqThoremProp}
\end{equation}
where $x_i=g_1(\tilde{x},\tilde{y})$ and $y_i=g_2(\tilde{x},\tilde{y})$ are the inverse mapping at M1 for the independent variables $(\tilde{x},\tilde{y})$ and $S_0 = S(g_1(\tilde{x},\tilde{y}),g_2(\tilde{x},\tilde{y}),\tilde{x},\tilde{y})$.
\end{theorem}

The proof of this theorem is  given in Appendix B. In the remaining sections we will use Eq.~\ref{EqThoremProp} to compute the diffraction through a square PIAA system.

\subsection{Validity of the second order approximation}
A direct consequence of Theorem~\ref{MainPropThorem} is a new  derivation of Eq.~\ref{EqRemap} in the ray optics limit.  If we consider the behavior of Eq.~\ref{EqThoremProp} as $\lambda \rightarrow 0$ or equivalently $D \rightarrow \infty$, then the asymptotic value of the complex Gaussian in Eq.~\ref{EqThoremProp} is the determinant of the Jacobian of the change of variables defined by the pupil remapping. Such a result can be rigorously proven using the stationary phases approximation (\cite{BornAndWolf}). We thus find that
\begin{equation}
E_{out}(\tilde{x},\tilde{y}) = \frac{1}{\sqrt{\frac{\partial f_1} 
{\partial x} \frac{\partial f_2}{\partial y} - \frac{\partial f_2} 
{\partial x} \frac{\partial f_1}{\partial y}}} = \sqrt{\frac{\partial  
g_1}{\partial \tilde{x}} \frac{\partial g_2}{\partial \tilde{y}} -  
\frac{\partial g_1}{\partial \tilde{y}}\frac{\partial g_2}{\partial  
\tilde{x}}}.
\label{eq:stationary_phases}
\end{equation}
The goal of the pupil mapping design is to set the field at the exit  
pupil equal to the desired apodization, $A(\tilde{x},\tilde{y})$. Equating  Eq. \ref{eq:stationary_phases} to  $A(\tilde{x},\tilde{y})$ indeed leads to Eq. \ref{EqRemap}, the Monge-Ampere equation. This analytical result not only provides a sanity  
check that our diffractive optics approach asymptotically behaves as  
ray optics but also provides a unified theoretical understanding of  
pupil mapping systems. 
%

We now expand our discussion about  the validity of this second order approximation at  visible wavelength and with finite pupil sizes. The functional form of  the integral in Eq.~\ref{EqThoremProp} is similar to the Fresnel  approximation: it is composed of a quadratic exponential centered at  the location of the incident ray optics wavelets, $(x_i,y_i)$. The  corresponding effective propagation distance  has been adjusted  according to the effective focal length associated with the  corresponding ray ,$\dpart{f_1}{x}\big |_{x_i,y_i}$. We follow the presentation on pp 66 - 68 of \cite{Goodman} and adapt it to Eq.~\ref{EqThoremProp}. We are interested in proving that the terms of order higher than $2$ in the Taylor expansion of the Optical Path Length do not contribute to the value of the propagation integral. For a given point $(\tilde{x}, \tilde{y})$ at $M2$, these terms scale as:
\begin{eqnarray}
&\;&max_{M1}\{\frac{2 \pi}{\lambda}[Q(x,y,\tilde{x},\tilde{y}) -  Q_0 -  G_{2}(x,y,\tilde{x},\tilde{y}) ] \} \nonumber \\
&\propto& \frac{\pi}{4 \lambda Q_0^3}  max_{\mathcal{A}_{(x_i,y_i)}} \{ \Big ( \dpart{f_1}{x}\big |_{x_i,y_i}(x-x_i)^{2} + 2 \dpart{f_2}{x}\big |_{x_i,y_i}(x-x_i)(y-y_i) +  \dpart{f_2}{y}\big | _{x_i,y_i} (y-y_i)^{2} \Big )^2 \} \nonumber
\end{eqnarray}
where $\mathcal{A}_{(x_i,y_i)}$ stands for the region of $M1$,  centered around $(x_i,y_i)$, that mostly contributes to the value of the field at $M2$.  Following the local stationary phase applied to complex gaussians described in \cite{Goodman}, we find that the area of $\mathcal{A}_{(x_i,y_i)}$ is:
\begin{equation}
\frac{16 \lambda Q_0 }{{\sqrt{(\frac{\partial f_1}{\partial x}  \frac{\partial f_2}{\partial y} - \frac{\partial f_2}{\partial x}  \frac{\partial f_1}{\partial y}) \Big |_{x_i,y_i}}}}.
\end{equation}
Thus the condition on the higher order terms becomes:
\begin{equation}
max_{M1}\{\frac{2 \pi}{\lambda}[Q(x,y,\tilde{x},\tilde{y}) -  Q_0 -  G_{2}(x,y,\tilde{x},\tilde{y}) ] \} \simeq 16 \pi \lambda \ll 1 
\end{equation}

This condition is always true for the PIAA designs considered here. This discussion illustrates that the propagation integral through pupil re-mapping aspherical optics presented here is, from an analytical point of view,  just as accurate as the Fresnel integral when propagating through parabolic optics. Since for the remainder of this paper we will only focus on analytical evaluations of Eq.~\ref{EqThoremProp}, discussing its numerical precision and performance when integrated using a set of discrete two  dimensional  arrays is beyond our scope. This problem has been addressed in a separate communication \cite{pueyo:74400E}.  Moreover we recently  published a comprehensive comparison, \cite{krist:77314N}, between the numerical  accuracy of the present method and a S-Huygens model, \cite{2006ApJ...636..528V,2006ApJ...652..833B}. There we found that using a combination of both methods allowed large gains in computation speed and memory allocation at a small cost in numerical precision.

\subsection{Numerical propagator for square apertures}\label{NumProp}
We can now use theorem \ref{MainPropThorem}  to compute the  diffraction limited PSF for PIAA systems over square separable  apertures.  Because of the separability,  the cross term in Eq.~\ref{EqThoremProp} vanishes and we find
\begin{eqnarray}
E_{out}(\tilde{x},\tilde{y})= \frac{e^{\frac{2 i \pi}{\lambda} Q_0}}{2 i \lambda Z} \int \int e^{\frac{ i \pi }{\lambda S_0}  \big( \dpart{f_1}{x}\big |_{x_i,y_i}(x-x_i)^{2} +  \dpart{f_2}{y}\big |_{x_i,y_I} (y- y_i)^{2} \big) } dx dy.
\end{eqnarray}
If we drop the piston term in front of the integral and use Eqs.~\ref{EqXsep} and \ref{EqYSep} to substitute for the derivatives of the ray optics remapping we find
\begin{equation}
E_{out}(\tilde{x},\tilde{y})= \frac{1}{2 i \lambda Z} \int_{-D/2}^{D/2}  e^{\frac{ i \pi }{\lambda S_0 A_X(\tilde{x})^2}  (x-x_i(\tilde{x}))^{2} } dx \int_{-D/2}^{D/2} e^{\frac{ i \pi }{\lambda S_0 A_Y(\tilde{Y})^2}  (y-y_i(\tilde{y}))^{2} } dy.
\end{equation}
We then approximate $S_0 = Z$ and proceed to the following change of variables in the integrals:
\begin{eqnarray}
u &=& \sqrt{\frac{ 2 }{\lambda Z A_X(\tilde{x})^2} }(x-x_i(\tilde{x})) \\
v &= & \sqrt{ \frac{ 2 }{\lambda Z A_Y(\tilde{y})^2} } (y-y_i(\tilde{x}))
\end{eqnarray}
resulting in
\begin{eqnarray}
E_{out}(\tilde{x},\tilde{y}) = \frac{A_{X}(x_{i})A_{Y}(y_{i})}{4}  \int^{\sqrt{\frac{2}{\lambda Z}}\frac{x_{i}+\frac{D}{2}}{A_{X}(\tilde{x})}}_{\sqrt{\frac{2}{\lambda Z}}\frac{x_{i}-\frac{D}{2}}{A_{X}(\tilde{x})}} e^{\frac{ i \pi }{2} u^{2}} du  \int^{\sqrt{\frac{2}{\lambda Z}}\frac{y_{i}+\frac{D}{2}}{A_{Y}(\tilde{y})}}_{\sqrt{\frac{2}{\lambda Z}}\frac{y_{i}-\frac{D}{2}}{A_{Y}(\tilde{y})}} e^{\frac{ i \pi }{2} v^{2}} dv.
\end{eqnarray}
%
%
This expression can be easily calculated using the Fresnel Integrals since
\begin{equation}
\int^{b}_{a}e^{\frac{i \pi}{2}u^{2}} du = FresnelC(b)-FresnelC(a)+ i \; \big(FresnelS(b)-FresnelS(a)\big) .
\end{equation}

Thus computing $E_{out}(\tilde{x},\tilde{y})$ reduces to functional evaluations of the Fresnel integrals, which are tabulated in many computational packages,
\begin{eqnarray}
\mbox{FresnelC}(x) = \int_{0}^{x} \cos( \frac{\pi}{2} u^2) d u \\
\mbox{FresnelS}(x) = \int_{0}^{x} \sin(\frac{\pi}{2} u^2) d u.
\end{eqnarray}

\begin{figure}[htbf]
\begin{center}
\includegraphics[width=3in,angle=-90]{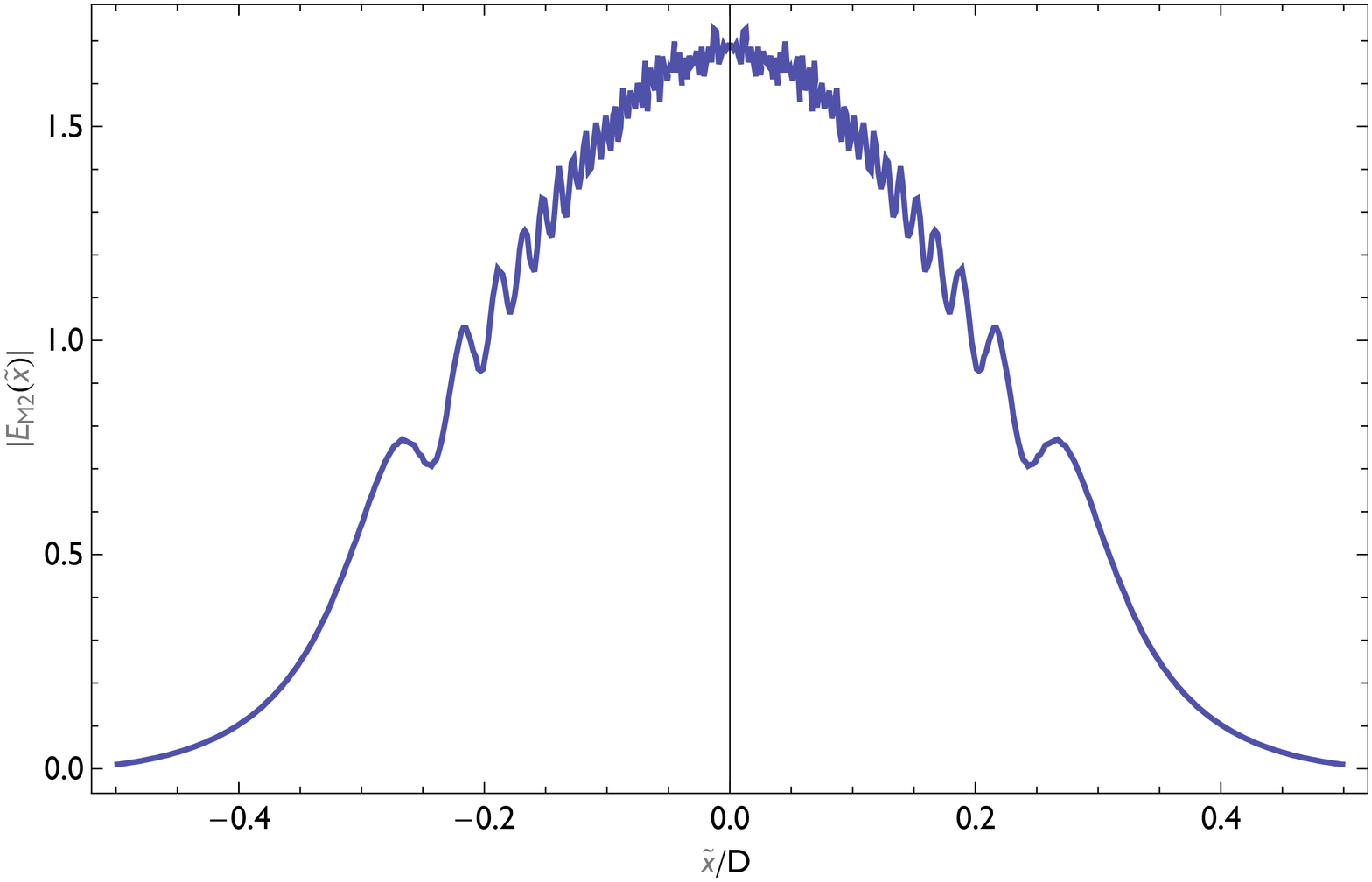}
\includegraphics[width=3in,angle=-90]{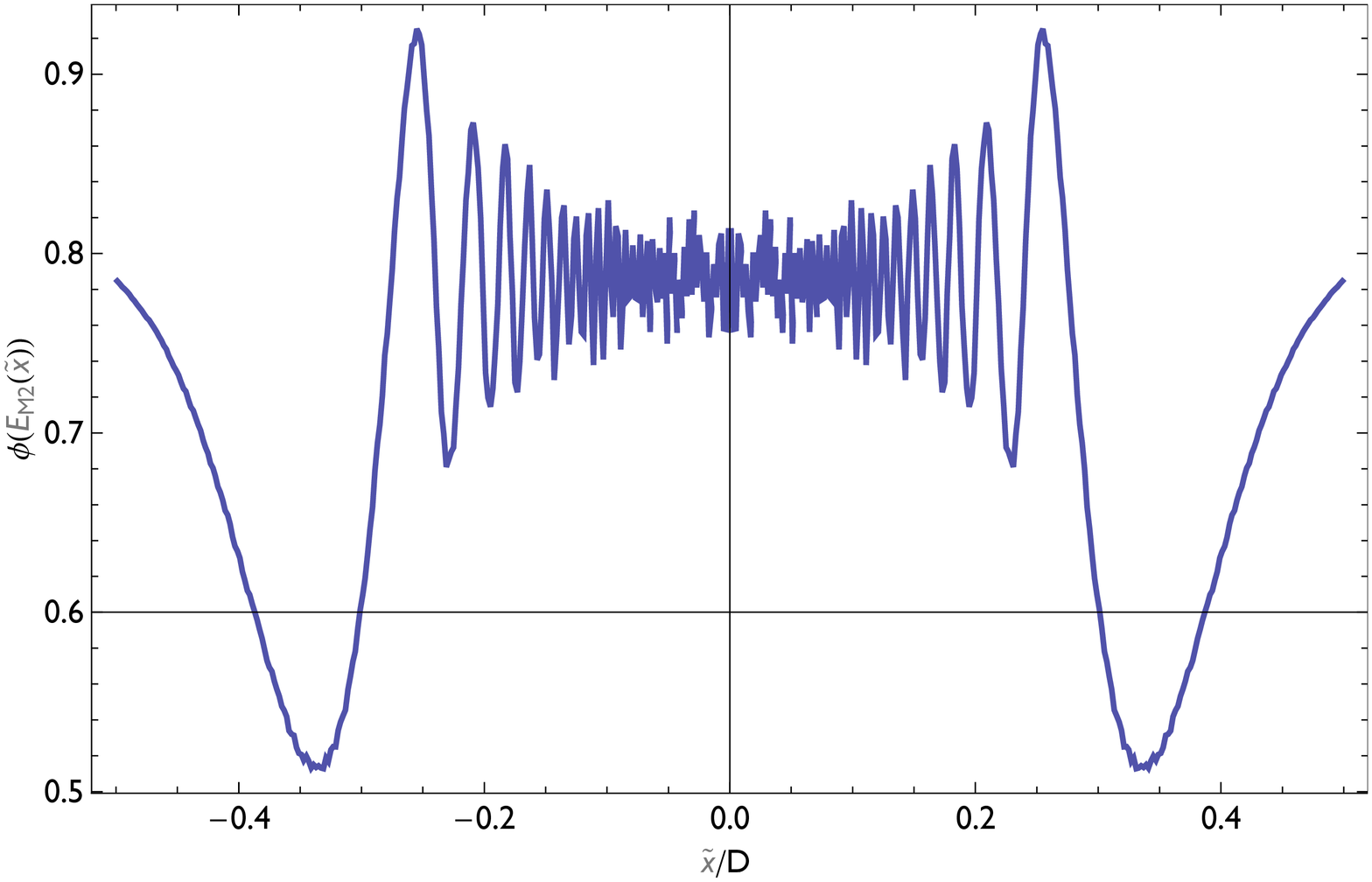}
\caption[Propagated field after $M2$ for a pure PIAA unit]{Propagated field after $M2$ for a pure PIAA unit. Top: Amplitude. Bottom: Phase. The x axis is in units of $\lambda /D _{On Sky}$. Aperture size $D= 3$ cm, mirrors distance $z=1$ m}
\label{Field1DpropNoApod}
\end{center}
\end{figure}

\begin{figure}[htbf]
\begin{center}
\includegraphics[width=3.5in,angle=90]{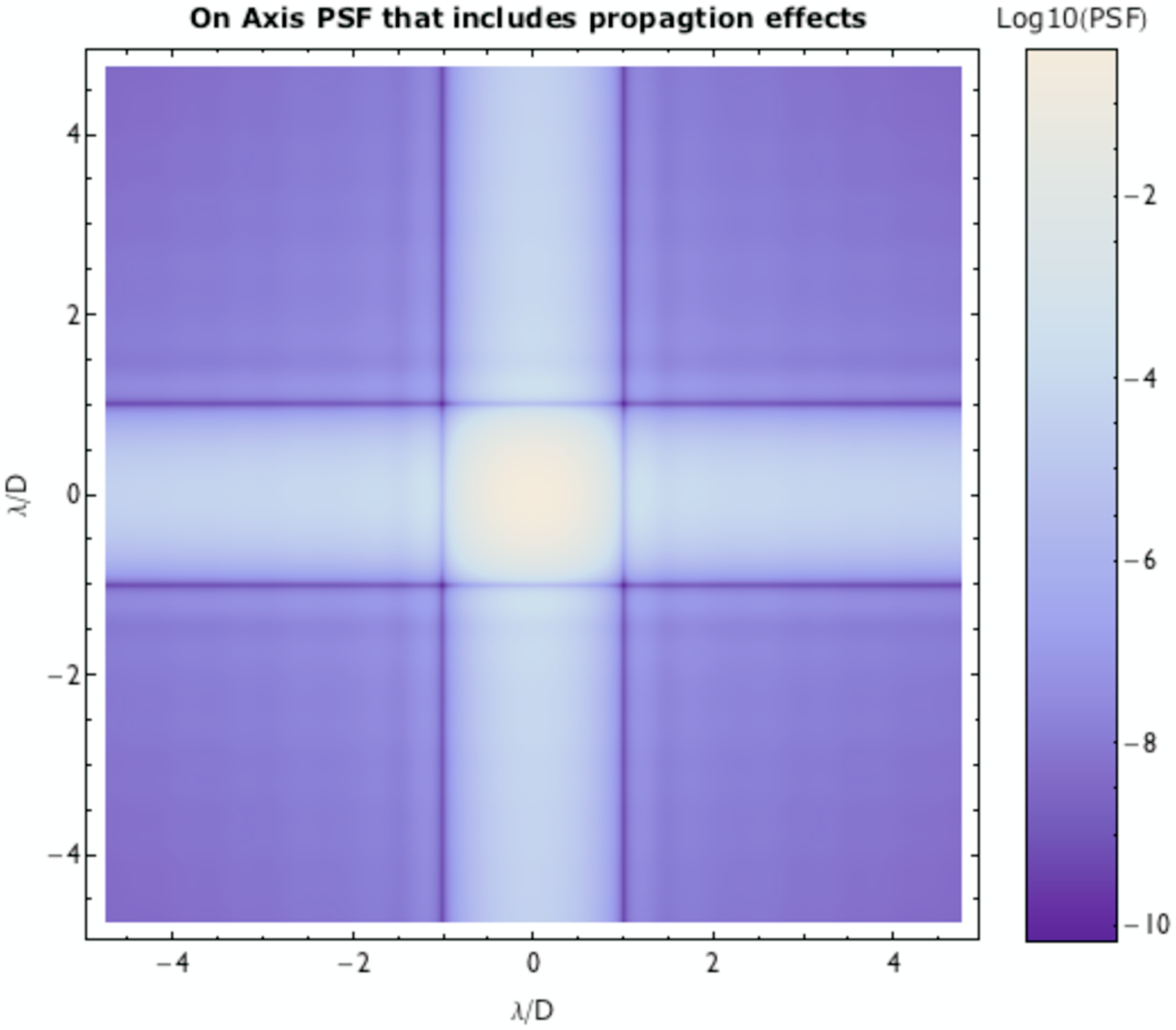}
\includegraphics[width=3.5in,angle=-90]{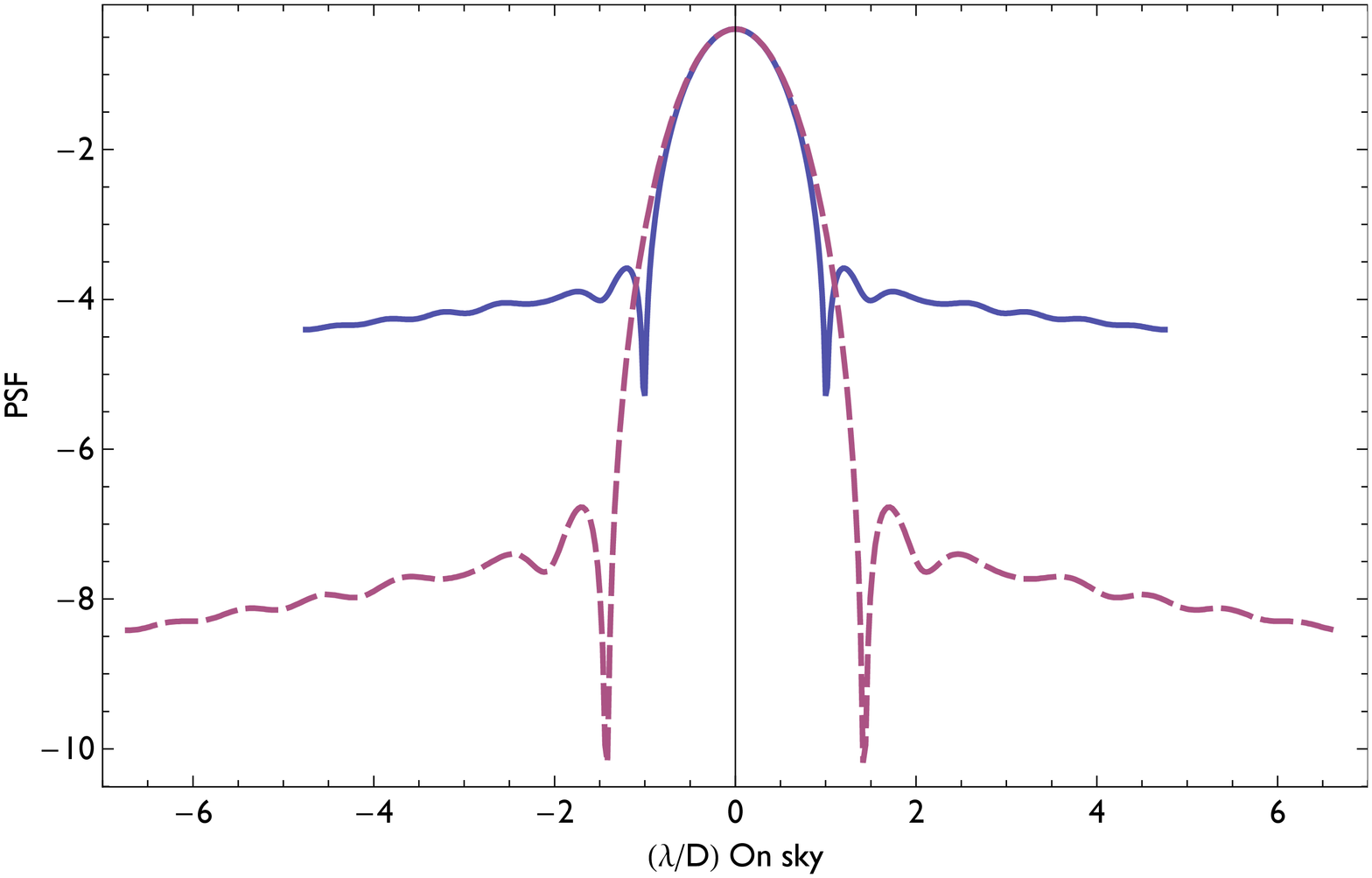}
\caption[Diffraction limited PSF]{Diffraction limited PSF. Top: Two dimensional PSF. Bottom: One dimensional profiles along the horizontal axis and the diagonal of the PSF. The x axis is in units of $\lambda /D _{On Sky}$. Aperture size $D= 3$ cm, mirrors distance $z=1$ m}
\label{PSFpropNoApod}
\end{center}
\end{figure}

Fig.~\ref{Field1DpropNoApod} shows the field after $M2$ that has been computed using this propagator. Diffraction effects result in field oscillations that degrade the contrast. The PSF, for square mirrors of size $D=3$ cm separated by $z = 1$ m, is shown in Fig.~\ref{PSFpropNoApod}, where the diffraction limited contrast is $10^{-8}$. Note, however, that for a separable square geometry the contrast floor due to edge propagation effects, with perfect optics, no wavefront error and no apodiser,  is already at the $10^{-8}$ level, as compared to the $10^{-5}$ floor in circular geometries. Since we chose to use a separable amplitude profile on the square aperture, with a soft target apodisation along each dimension, we can take advantage of the deeper contrast property along the diagonal. As a consequence the limit on contrast induced by edge propagation effects is lower than for circular apertures. This contrast is sufficient enough for the detection of self-luminous Jovian planets, but edge effects need to be mitigated in order to enable the detection of exo-planets.  Note that for moderate contrast there also exist azimuthally symmetric solutions that do yield the required contrast to image self-luminous planets in the infra-red and only require one of the apodizers presented in the next section. In particular \cite{2005astro.ph.12421P} showed that a system designed according to geometrical optics only, but including an apodizer to allow manufacturability of the PIAA optics, can be used in conjunction with a single deformable mirror in order to yield a $10^{7}$ contrast in a $20$ percent bandwidth. 

\section{Hybrid Apodisers-PIAA Designs on Square Apertures}\label{sec:Hybrid}
\subsection{Pre- and post-apodisers}

\begin{figure}
\begin{center}
\includegraphics[width=3.7in]{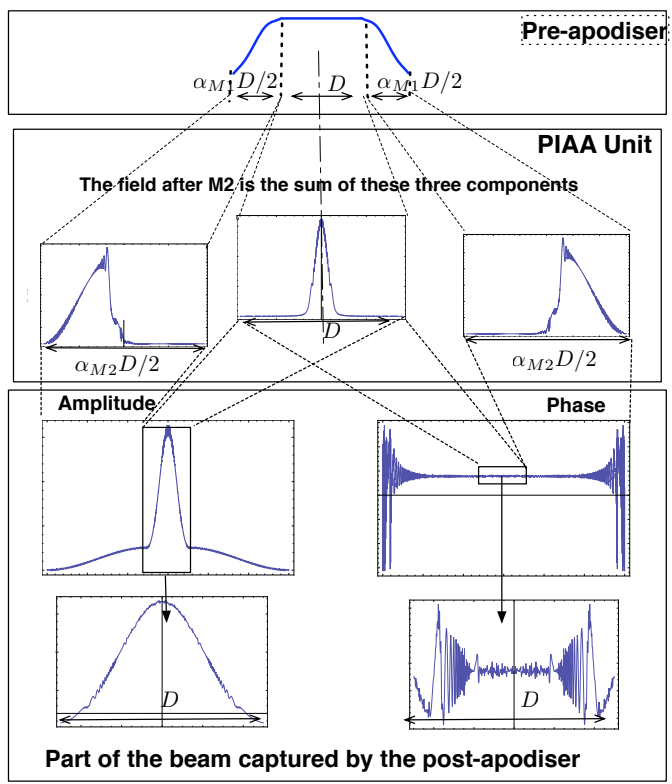}
\end{center}
\caption[Actual values of the field across the PIAA unit]{Electrical field propagated through the PIAA unit. Top panel: transmissivity profile of the pre-apodiser. Note that the pupil needs to be oversized by a factor $\alpha_{M1}$ in order to accommodate for the smooth roll off at it s edges. This results in a loss of throughput.  Middle panel: effect of the propagation through the PIAA aspherical optics; most of the light is concentrated in the center of the beam, the smooth edges of the pre-apodiser help damp the edge diffraction ripples. Note that these ripples are spatially extended over  large zones of dimensions $\alpha_{M2} D$ surrounding the transmissive area of the post-apodiser. Bottom panel: the post-apodiser retrieves the expected prolate profile and clips the high frequency edge ripples out of the main beam path of the coronagraph. Note that the phase oscillations actually transmitted are much smaller that the ones in the outer region that has been suppressed by the post-apodiser. While in general $\alpha_{M2} \ggg \alpha_{M1}$ their values have been exaggerated in this figure for illustration purposes}
\label{FigCartonnRealValues}
\end{figure}

In order to design a pupil remapping system that will create a broadband $10^{-10}$ null in the image plane of the telescope,  \cite{2005astro.ph.12421P} introduced a set of pre- and post-apodisers to smooth the edges of the pupil and reduce the diffraction effects. We revisit the roles of these apodising screens in light of Theorem~\ref{MainPropThorem}.  These screens have two effects:
\begin{itemize}

\item a)\textit{Suppress the field oscillations due to diffraction from the edges of $M1$}. The discontinuity at the edges of $M1$ is softened by slightly oversizing the incoming beam and applying a smooth apodisation to the oversized portion of the pupil. This soft edge needs to be mapped out of the main beam path in order to preserve the high contrast apodisation profile. As a consequence, the outer part of $M1$, that is oversized by a fraction $\alpha_{M1}$ of the pupil diameter, is designed to be parabolic and diverging.  This spreads the edge oscillations outside the edges of $M2$, resulting in a reduction of the magnitude of the phase oscillations on $M2$. The functional form of the pre-apodiser is completely arbitrary as long as it smoothly goes to zero  outside the edges of $M1$ (is continuous with continuous first derivative).  In practice, we choose a function to facilitate the calculation of the diffraction propagation using the special functions $FresnelC$ and $FresnelS$.  Most current PIAA designs use a simple cosine apodization,
%
\begin{equation}
A_{pre}(x)  = \left\{\begin{array}{clc}
\frac{1}{2} \left(1 + \cos [ \frac{2 \pi}{D \alpha_{M1}} (x+ \frac{D}{2}) ] \right)  & \mbox{for } & - \frac{ (1+\alpha_{M1}) D}{2}< x  < -\frac{D}{2}  \\
1 & \mbox{for} & -\frac{D}{2} < x < \frac{D}{2}   \\
\frac{1}{2} \left(1 + \cos [ \frac{2 \pi}{D \alpha_{M1}} (x- \frac{D}{2}) ]\right)  & \mbox{for} & \frac{D}{2}< x  < \frac{ (1+\alpha_{M1}) D}{2}\end{array}  \right. \label{Eq::PreApod}
\end{equation}

In Appendix C we show how this apodization results in semi-anlaytical solutions for the propagation integral.  Current research is being carried out in order to explore optimal choices for the pre-apodizer.

\item b) \textit{Limit the strength of the remapping}. By limiting the amount of apodization in the 
ray optics design, the diffractive effects at M2 due to the strong remapping are reduced. In practice this is obtained by using a PIAA unit whose net apodisation on the edges of the pupil is weaker. This means introducing a saturation at the edges of the prescribed apodisation so that its net value never goes below a pre-determined value that we call $\beta$. The cancellation of the field at the edges of $M2$ is then achieved using a post apodiser that allows retrieval of the ideal prolate profile $A_X(\tilde{x})$. Again, how we relax the apodization and the post-apodizer we use to recover it is arbitrary.  Here, we use the simple functions suggested by  \cite{2005astro.ph.12421P},
\begin{eqnarray}
A_{PIAA}(\tilde{x}) &=&\eta \frac{A_X(\tilde{x}) + \beta}{1+\beta}  \nonumber \\
A_{post}(\tilde{x}) &=& \frac{A_X(\tilde{x})}{A_{PIAA}(\tilde{x})}
 \label{Eq::PostApod}
\end{eqnarray}
where $\eta$ is a normalization parameter that ensures that $A_{PIAA}(\tilde{x}) ^2$ integrates to unity so that area is preseved within the non-oversized portion of the beam. The two free parameters, $\alpha_{M1}$ and $\beta$ define the choice of apodising screens. Note that the strip of width $\alpha_{M1} D$ at the edge of $M1$ is remapped into a strip of width $\alpha_{M2} D$ at $M2$, and clipped out of the optical train, as shown on Fig.~\ref{FigCartonnRealValues}. 
\end{itemize}

The relationship between $\alpha_{M2}$ and $\eta$ and the two free parameters $\alpha_{M1}$ and $\beta$ can be derived as follows. First, the scaling $\eta$ of the prescribed PIAA apodisation is chosen such that the area within the actual optical train of the coronagraph is conserved:
\begin{eqnarray}
D &=& \int_{-D/2}^{D/2} A_{PIAA}(\tilde{x})^2  d \tilde{x} \\
\eta^2 &=& \frac{1}{\int_{-1/2}^{1/2} (\frac{A_X(u/D) + \beta}{1+\beta}  )^2  d u}
\end{eqnarray}
Because we seek a continuous curvature for the surface of M1 we also impose the value of the prescribed PIAA apodisation for $|\tilde{x}| > D/2$ to be such that:
\begin{equation}
A_{PIAA}(\tilde{x}) =\eta \frac{A_X(D/2) + \beta}{1+\beta} \simeq \eta \frac{ \beta}{1+\beta} 
\end{equation} 
since $A_X(\tilde{x})$ is almost zero near to the edge of the pupil. Then, using energy conservation, the strip of width $\alpha_{M1} D /2$ is now remapped at M2 into a strip of width 
\begin{equation}
\alpha_{M2} D = \frac{\alpha_{M1}  D}{A_{PIAA}(D/2)^2 } = \alpha_{M1} (\eta \frac{ \beta}{1+\beta} )^{-2} D /2.
\end{equation}
This yields,
\begin{equation}
\alpha_{M1} = \alpha_{M2} A_{PIAA}(D/2)^2 =\alpha_{M2} (\eta \frac{ \beta}{1+\beta} )^2.
\label{Eq::remapstrip}
\end{equation}
Thus, since $\beta << 1$ and $\eta \simeq 1$, the thin strip at the edge of $M1$ is remapped into a thick area at the edges of $M2$: $\alpha_{M1} <<\alpha_{M2}$.  Note that for a given $\beta$,  Eq.~\ref{Eq::remapstrip} corresponds to a one to one mapping between $\alpha_{M1}$ and $\alpha_{M2}$. Thus we do not loose generality when, for practical reasons, we choose to use $\alpha_{M2}$ as the free design parameter.

In the next section we present several square PIAA designs. To do so we first define the post-apodiser by choosing $\alpha_{M2}$ and $\beta$ and then use Eq.~\ref{Eq::remapstrip} to compute $\alpha_{M1}$ and thus design the pre-apodiser.  

We derived here the relationships between $\alpha_{M2}$ and $\eta$ and $\alpha_{M1}$ and $\beta$  under the assumption that the pupil mapping optics conserve the area within the apodization region seen by the coronagraphs. This implies that the actual physical size of $M1$ is $(1+\alpha_{M1}) D$ and only its central portion of width $D$ is transmitted to the corononagraph. One might want to choose a different convention and use $D$ as the physical size of $M1$ and scale all the quantities accordingly. In that case $M2$ and the transmissive region of the post- apodiser are smaller than $D$. We summarize both conventions in Table \ref{Tab::PIAAConv}. In the present paper we chose to use the convention in the first line of Table \ref{Tab::PIAAConv}. Finally, we  point out that for large pupil sizes, whose edge diffraction effects are relatively small, one can design a PIAA unit with $\alpha_{M1} > 0 $ without using a pre-apodiser to taper the edges of $M1$. However in an effort to keep the scope of this paper as broad as possible, we present designs that  use pre-apodisers.  
%

\subsection{Numerical results and performance}
\begin{figure}[htbf]
\begin{center}
\begin{tabular}{cc}
\includegraphics[width=2.5in,angle=-90]{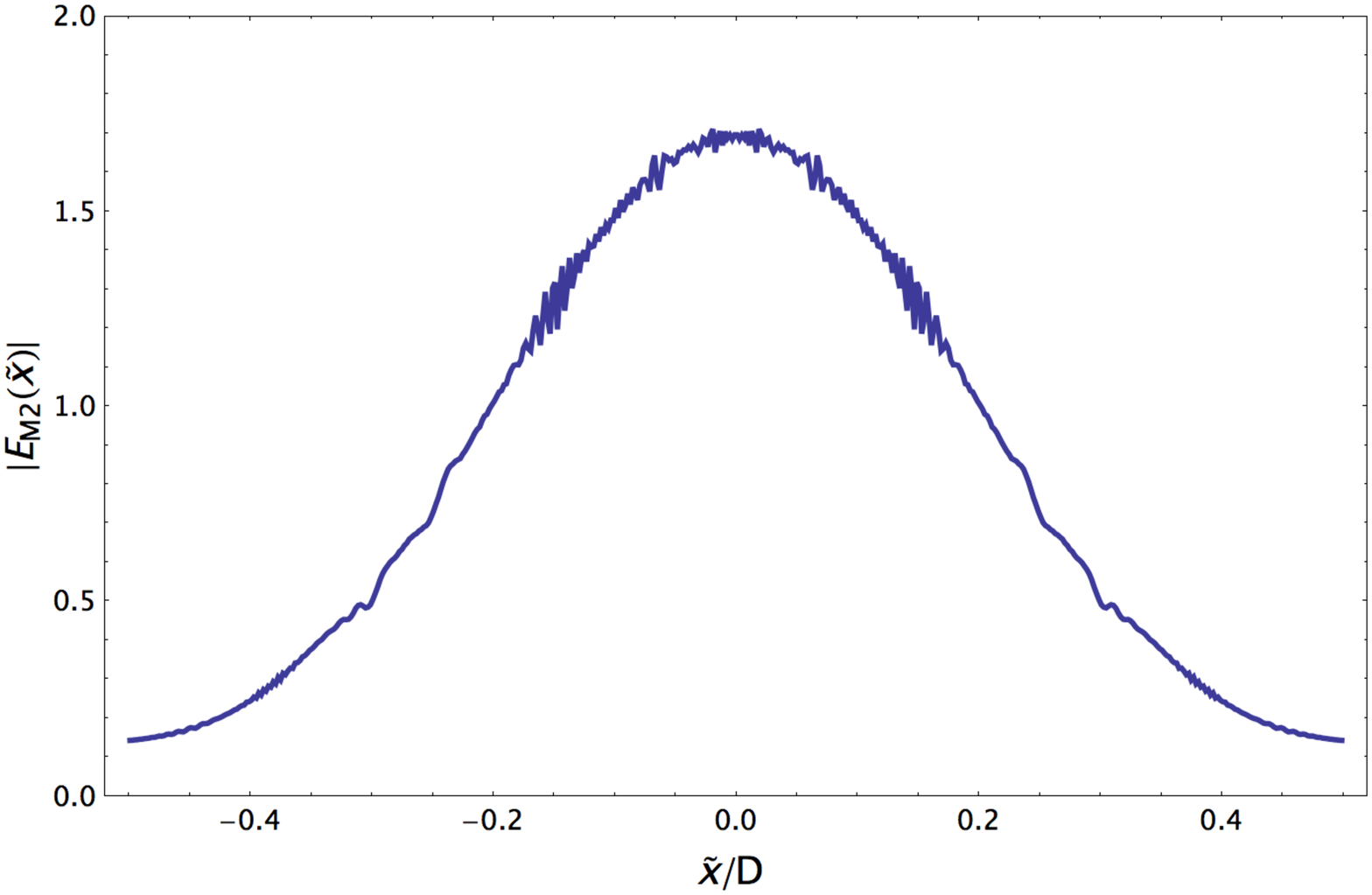} & \includegraphics[width=2.5in,angle=-90]{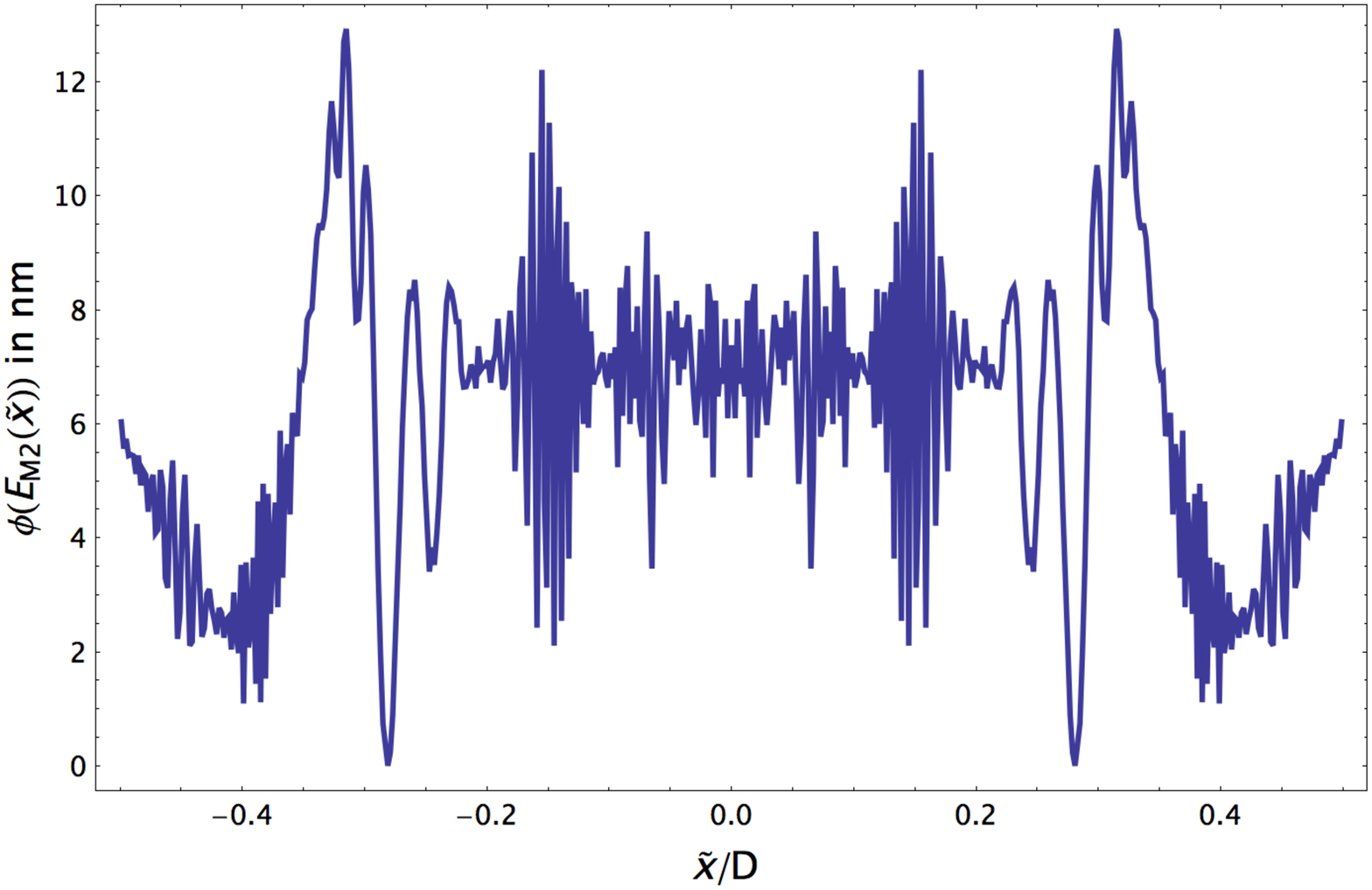}\\
\multicolumn{2}{c}{\includegraphics[width=5in,angle=-90]{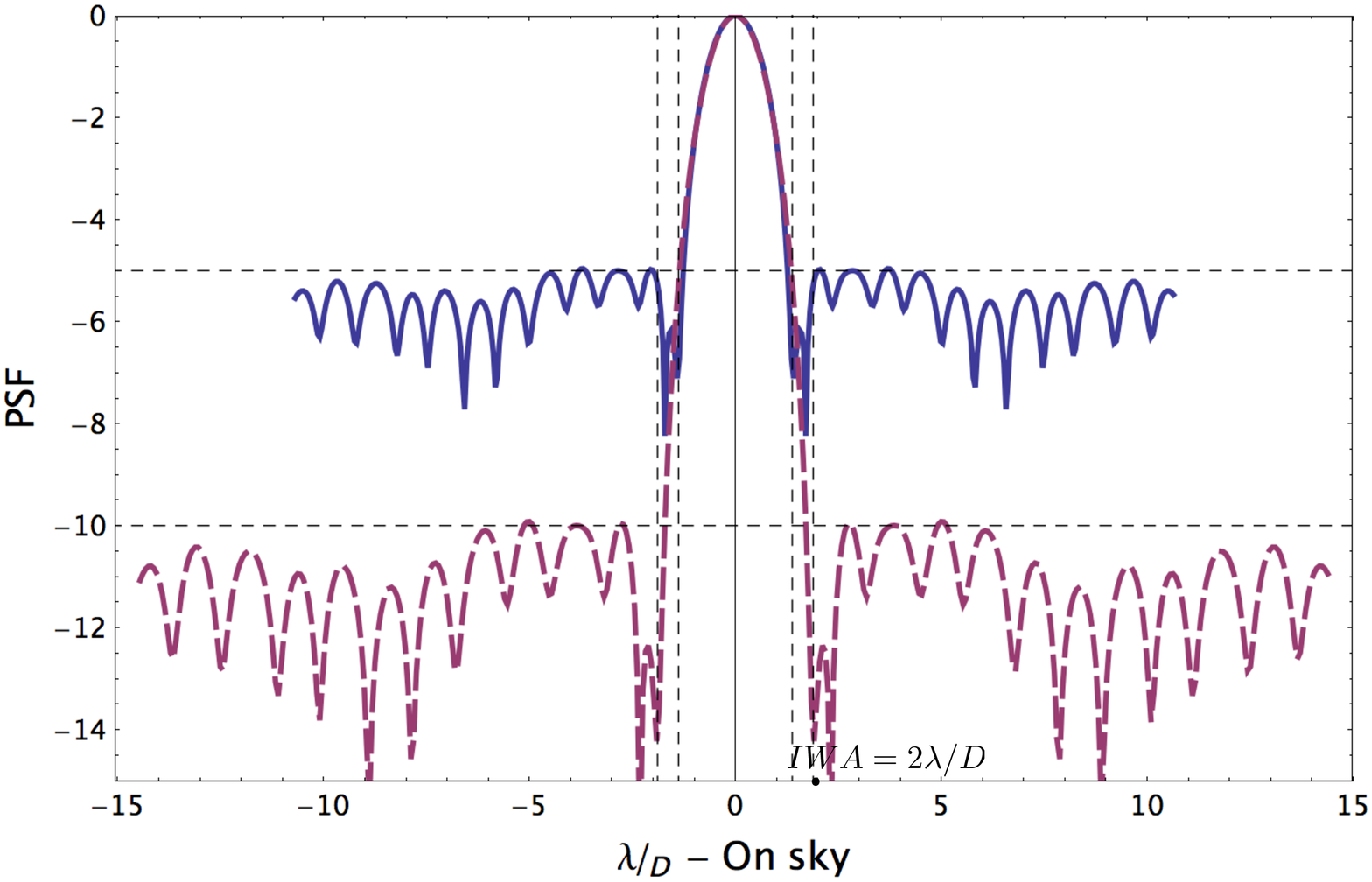}}
\end{tabular}
\end{center}
\caption[Propagated field after $M2$ for a hybrid PIAA unit]{Top: Propagated field after $M2$ for a hybrid PIAA unit. Left: Amplitude. Right: Phase. Aperture size $D= 3$ cm, mirrors distance $z=1$ m. Bottom: PSF of a hybrid PIAA unit. The throughput of the design shown here is $0.55$.}
\label{Field1DpropPostApodZoom3cm}
\end{figure}

\begin{figure}[htbf]
\begin{center}
\begin{tabular}{cc}
\includegraphics[width=2.5in,angle=-90]{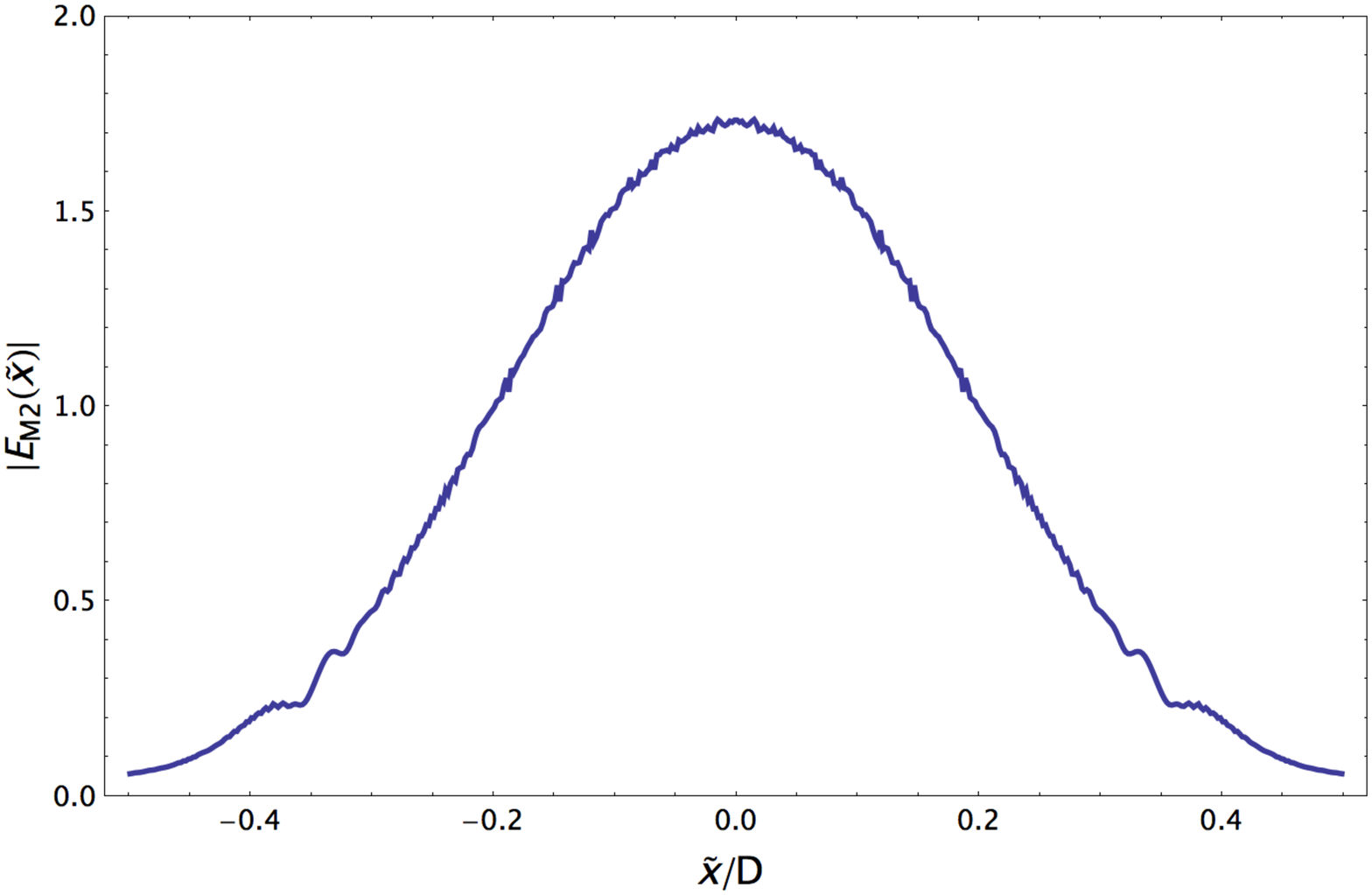} & \includegraphics[width=2.5in,angle=-90]{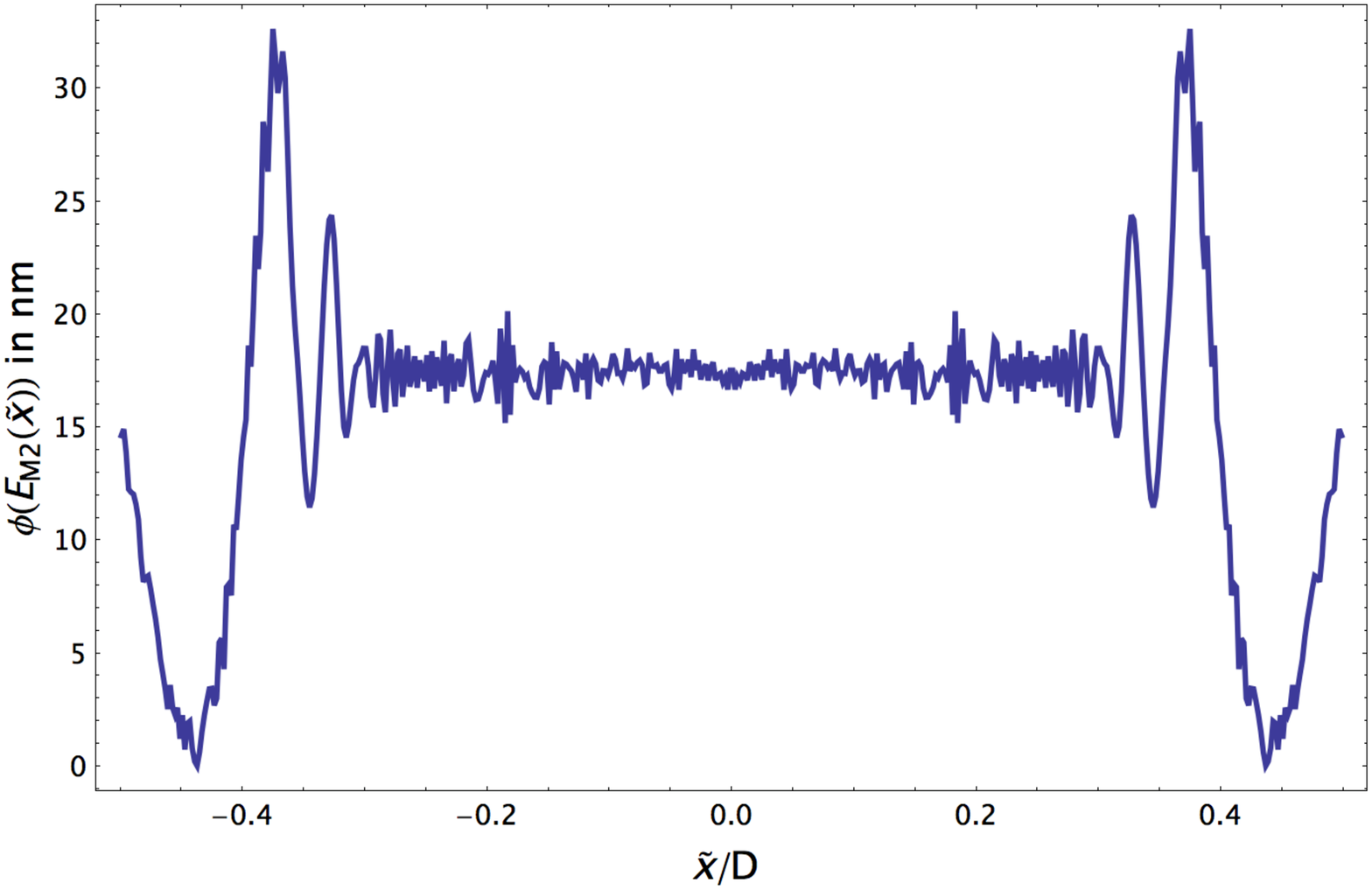}\\
\multicolumn{2}{c}{\includegraphics[width=5in,angle=-90]{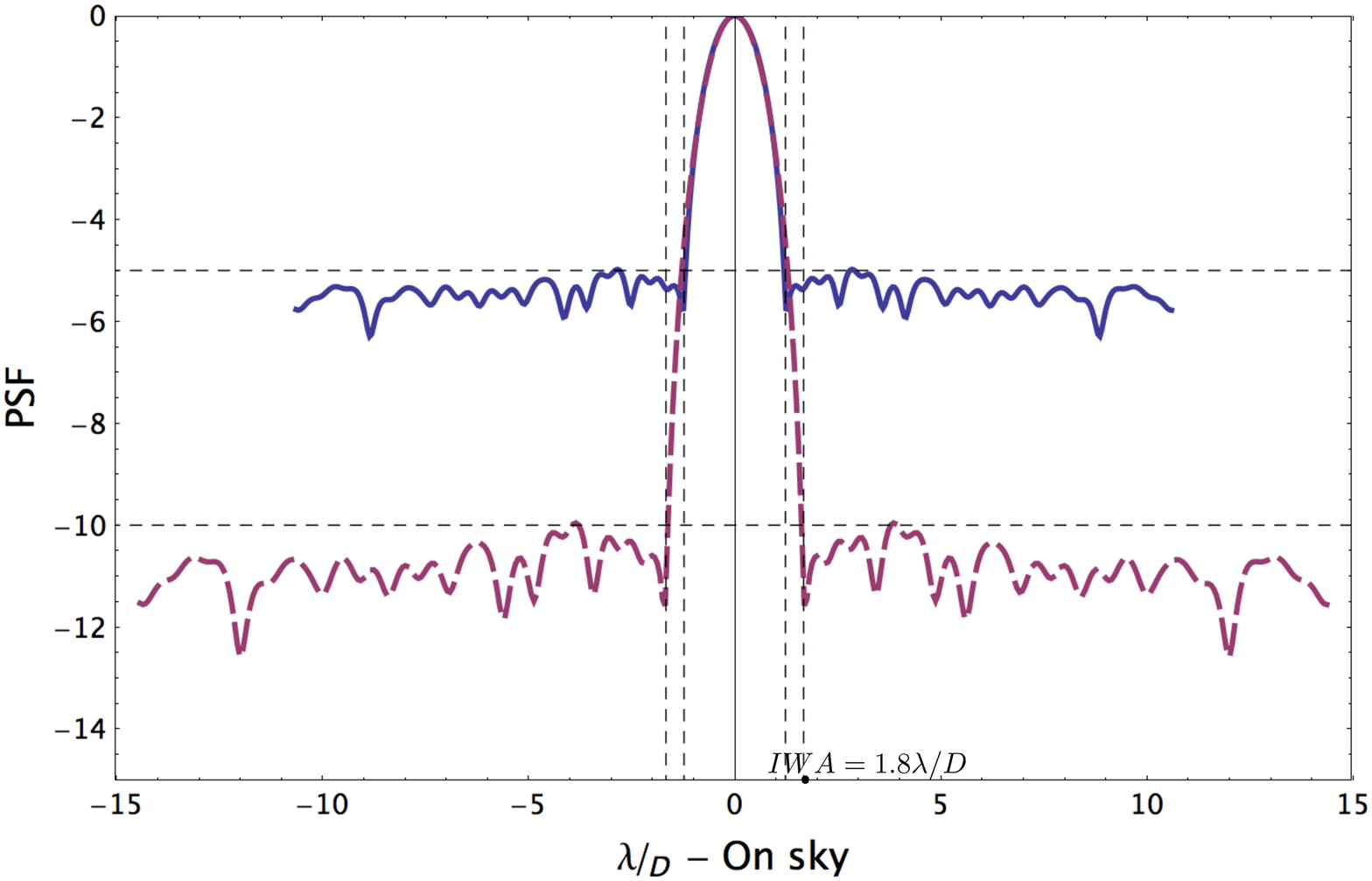}}
\end{tabular}
\end{center}
\caption[Propagated field after $M2$ for a hybrid PIAA unit]{Top: Propagated field after $M2$ for a hybrid PIAA unit. Left: Amplitude. Right: Phase. Aperture size $D= 9$ cm, mirrors distance $z=1$ m. Bottom: PSF of a hybrid PIAA unit.  The throughput of the design shown here is $0.89$.}
\label{Field1DpropPostApodZoom9cm}
\end{figure}

\begin{figure}[htbf]
\begin{center}
\includegraphics[width=2.4in,angle=-90]{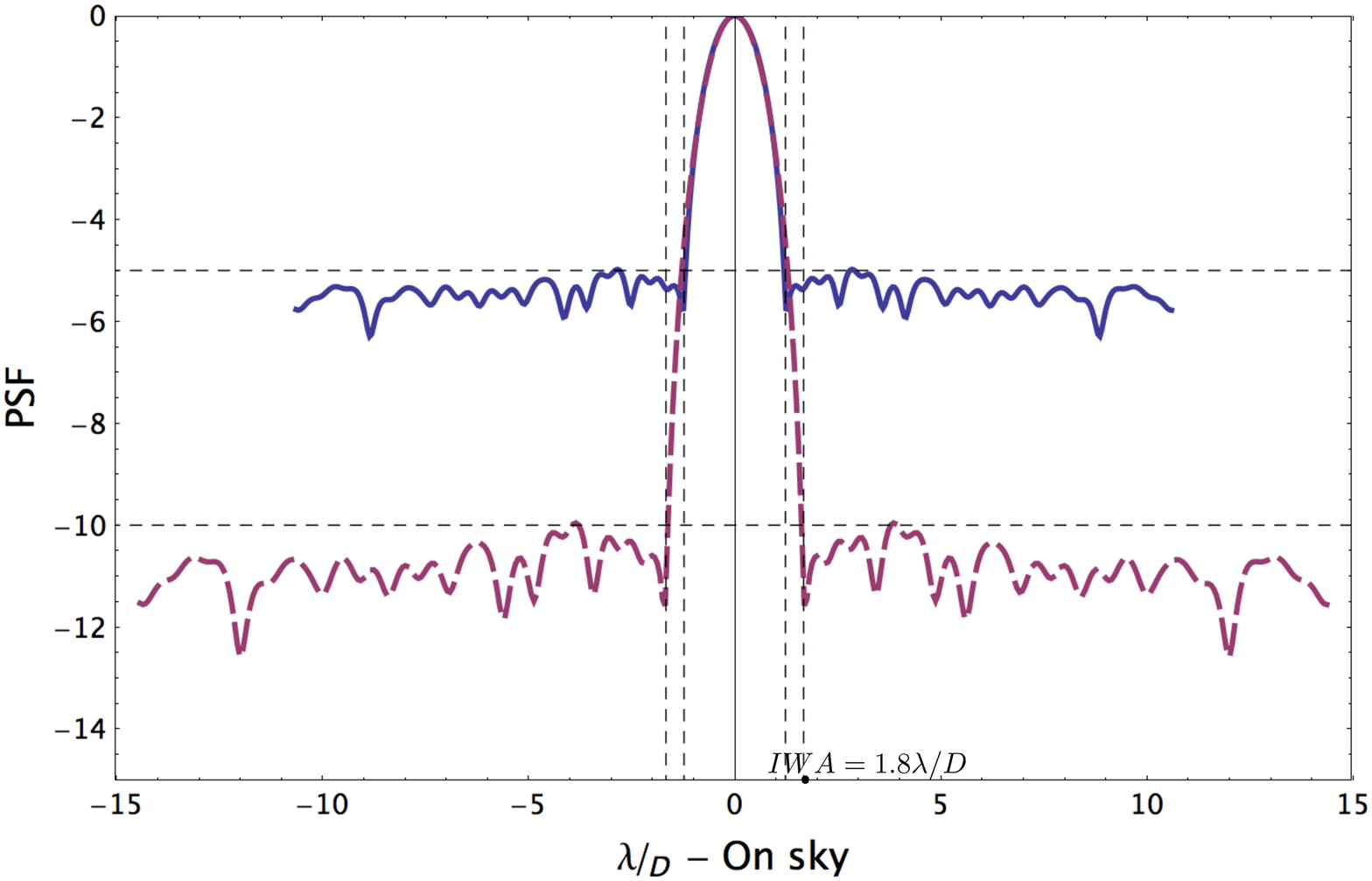} \\ \includegraphics[width=2.4in,angle=-90]{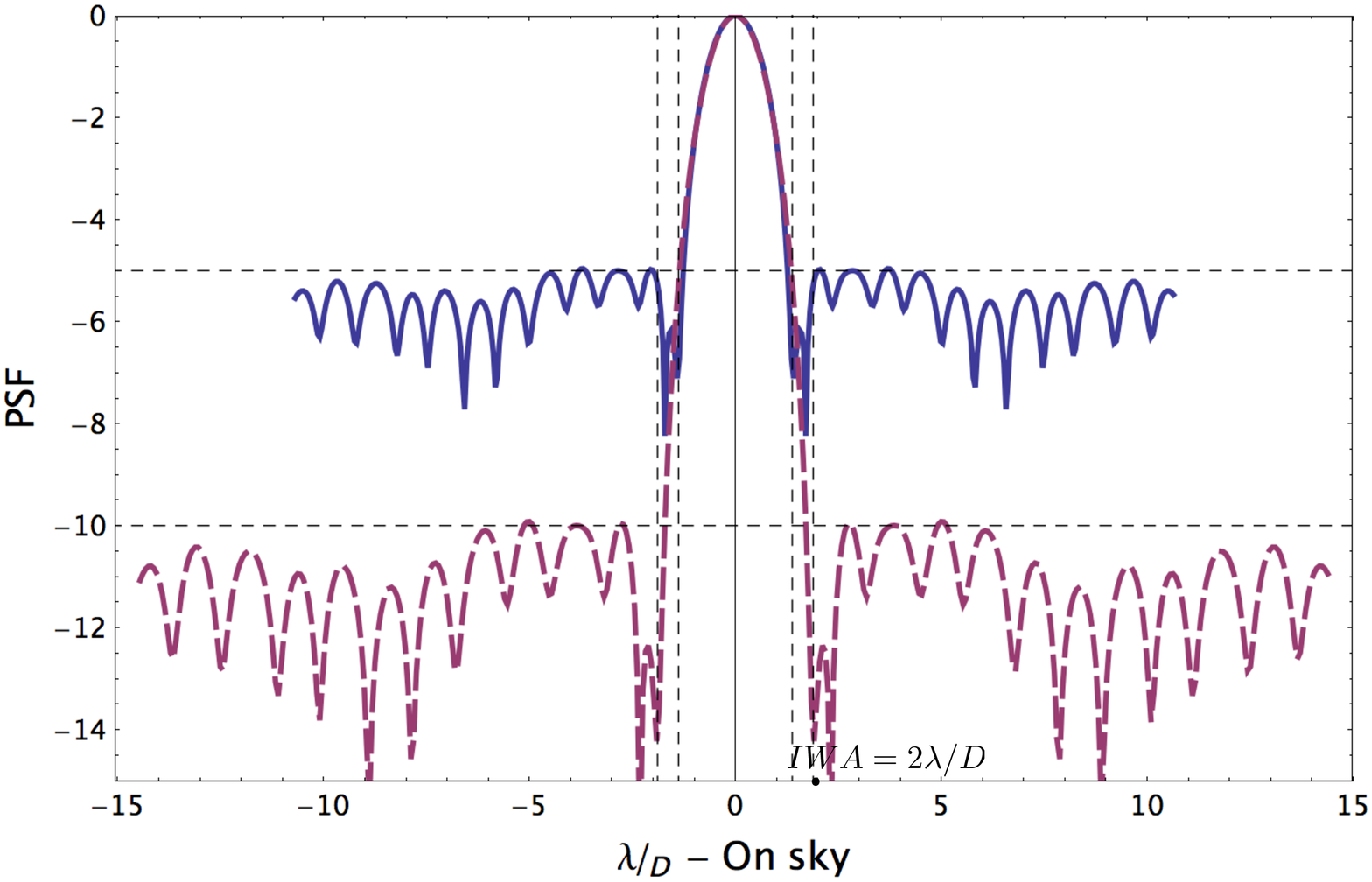} \\ \includegraphics[width=2.4in,angle=-90]{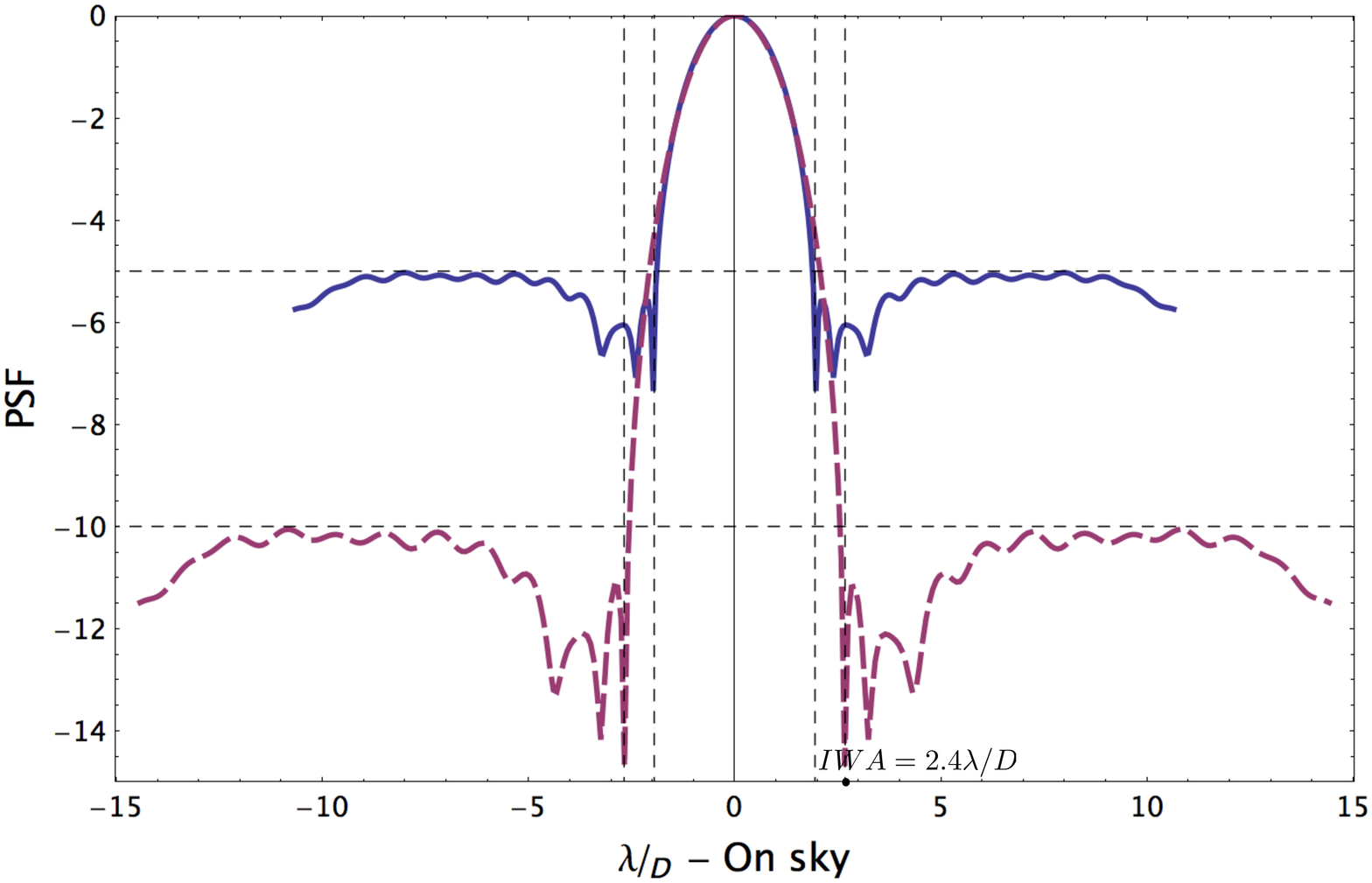}  \\
\end{center}
\caption[Performances of three different PIAA designs]{Monochromatic PSF of three different square PIAA designs. The pre-and post apodiser have been chosen so that the PSF features a $10^{-10}$ contrast along the diagonal. Top: $9$ cm mirrors, separated by $1$ m. Middle: $3$ cm mirrors, separated by $1$ m. Bottom: $1$ cm mirrors, separated by $1$ m. Note that as the pupil size decreases, the apodisers need to be stronger. This implies that the throughput decreases and the IWA increases.}
\label{Fig::ThreeMonoPSFs}
\end{figure}

Fig~\ref{FigCartonnRealValues} illustrates how the light propagates over a one dimensional aperture for such pre- and post-apodised designs. This figure shows true electrical field distributions computed using our analytical propagator. For the designs in Fig.~\ref{Field1DpropPostApodZoom3cm} and \ref{Field1DpropPostApodZoom9cm}  we respectively chose $\alpha_{M2}  = 4$ and $\beta = 0.15$, which yielded $\alpha_{M1} =  0.08$ and a stroke of 62$\; \mu m$, and $\alpha_{M2}  = 4$ and $\beta = 0.05$ which yielded $\alpha_{M1} =  0.01$ and a stroke of 585$\; \mu m$. The large edge oscillations are now remapped very far from the optical axis and thus have little impact on the contrast. Combined with the fact that M2 is only reflective for $|\tilde{x}|< D/2$, this effect has a highly beneficial impact on contrast: the edge oscillations are not captured by $M2$ at a very small cost in throughput. Because of the trigonometric form we chose for the pre-apodiser in this study, predicting the contrast and performances of such PIAA designs can be achieved without any numerical integration, using only well established functional evaluations. In Figs~\ref{Field1DpropPostApodZoom3cm} and \ref{Field1DpropPostApodZoom9cm} we show the field at $M2$ for pre- and post-apodised square PIAA coronagraphs with respective pupil sizes of $3$ cm and $9$ cm. In both cases, the phase oscillations at $M2$ are small enough so that the PSFs exhibit a $10^{-10}$ contrast on the diagonal axis  (bottom panels of  Figs~\ref{Field1DpropPostApodZoom3cm} and \ref{Field1DpropPostApodZoom9cm}). For each geometry, that is, each choice of $D$ and $z$, there is an entire family  of apodising parameters $(\alpha_{M1}, \beta)$ that will yield a $10^{-10}$ contrast. This parameter space will be explored in a future communication. Fig.~\ref{Fig::ThreeMonoPSFs} and Table~\ref{Tab::PIAAPerfs} illustrate three designs we obtained with pupil sizes $9$, $3$ and $1$ cm.

\begin{figure}
\begin{center}
\includegraphics[width=4in,angle=-90]{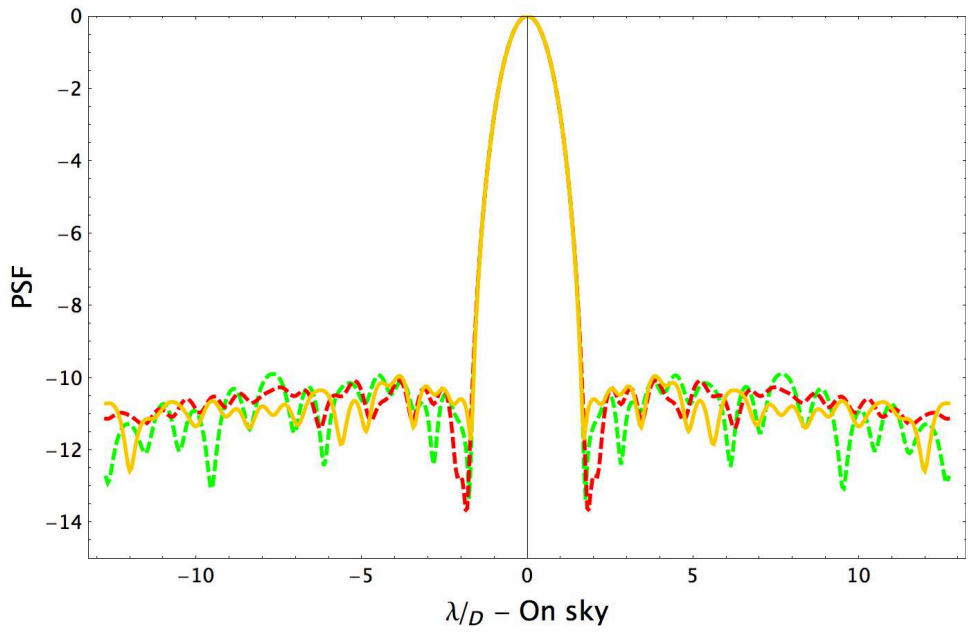}
\end{center}
\caption[Broadband contrast for]{Broadband contrast (540, 600 and 660 nm) for a hybrid PIAA solution with $9$ cm mirrors. This figure illustrates how the design presented on the top panel of Fig.~\ref{Fig::ThreeMonoPSFs} is truely a broadband design}
\label{FigPolyPSFs9cm}
\end{figure}

%
The contrasts of these design were found at a single wavelength of $\lambda = 600$ nm. To check the broadband performance, we computed the PSFs over a $120$ nm bandwidth (a  20$\%$ band centered on $\lambda$=600nm) with the same combination of $\alpha_{M2}$ and $\beta$ previously used. For this bandwidth, the $10^{-10}$ contrast is conserved, as shown on Fig.~\ref{FigPolyPSFs9cm}.  In general, the larger the wavelength considered, the larger the stellar halo is since propagation effects scale with $D^2/ (\lambda z)$.  A broadband optimization of PIAA will be presented in a subsequent communication.

\section{Feasibility of PIAA over square apertures: PIAA with deformable mirrors}
In this section we address some issues regarding the practical implementation of the PIAA solutions on square apertures.  Since most optical manufacturing tools are built in order to accommodate circular apertures, polishing a set of square ground PIAA mirrors such as the ones used to generate Fig.~\ref{FigPolyPSFs9cm}, with deformations close to $600 \; \mu$m, is a difficult endeavor. Herein we focus on the feasibility of a PIAA design using  square optics whose optical surfaces can be actuated with very high precision: Deformable Mirrors (DMs). Most of the current DMs envisioned for high contrast imaging applications are square, and are thus perfectly suited to implement the solutions discussed in this paper. The DM PIAA unit presented below was designed under the following constraints:

\begin{itemize}

\item The broadband contrast, $20$ percent bandwidth, should be below $10^{-10}$. 

\item The peak-to-valley deformation should be below $4 \; \mu$m for each DM. This number was chosen based on the state of the art maximal stroke for MEMS DMs.

\item The mirror curvature should be below the maximal curvature obtained by an actuator pushed at maximum stroke. In practice this maximal curvature constraint is already achieved by the use of a post-apodiser that enables a saturation of the DM induced apodisation.

\end{itemize}

The size of the active region of the DMs is fixed to $1$ cm which approximately corresponds to the size of the current large stroke MEMS DMs. We discuss here designs in agreement with the bottom panel of Fig.~\ref{Fig::ThreeMonoPSFs} and the bottom line of Table.~\ref{Tab::PIAAPerfs}. The free parameters are the separation $z$ between the DMs, the profile of the pre- and post-apodisers, and the DM deformation.

\begin{figure}
\begin{center}
\begin{tabular}{c}
\includegraphics[width=3.5in,angle=-90]{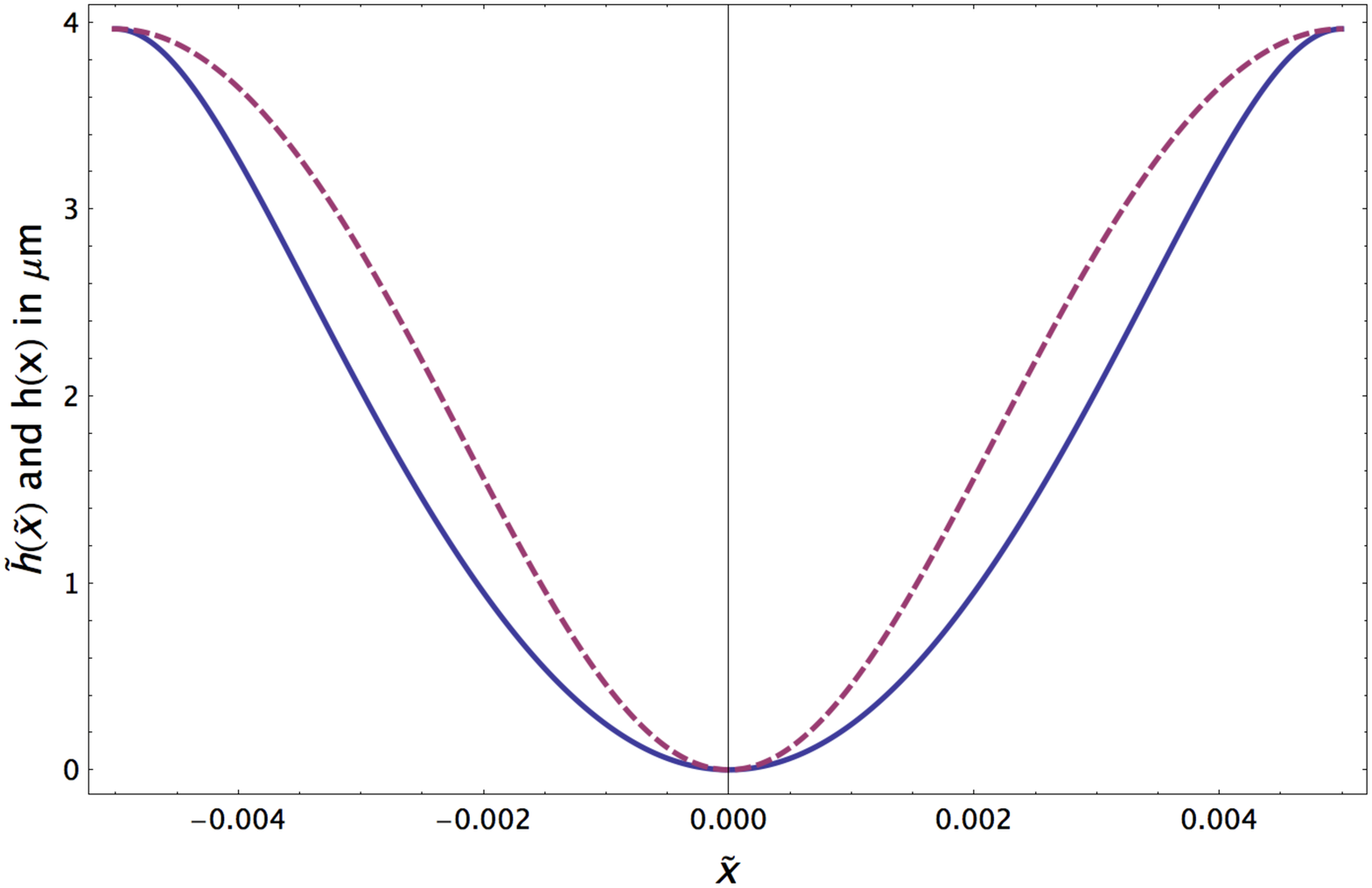} \\
\includegraphics[width=3.5in,angle=-90]{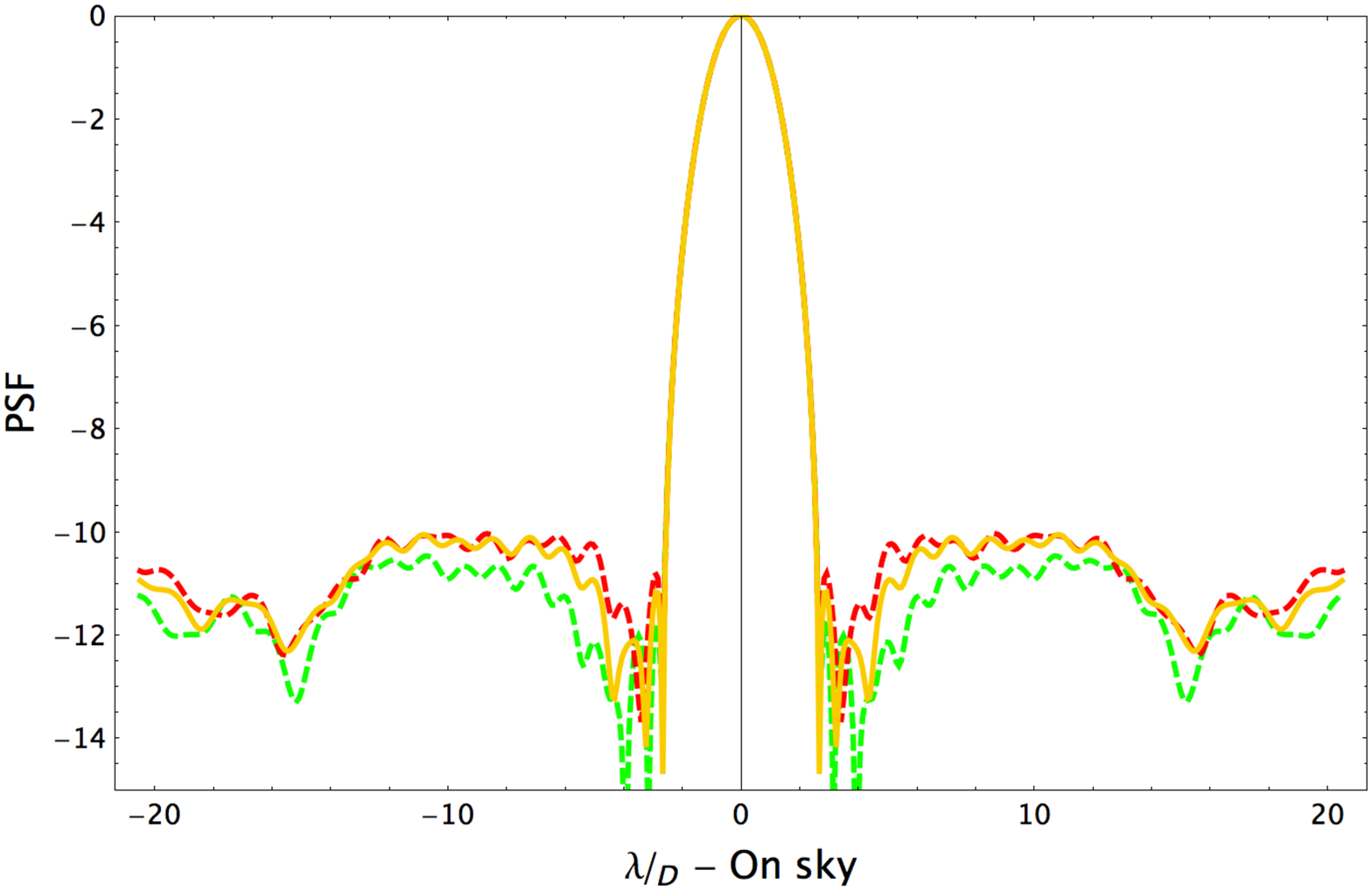}
\end{tabular}
\end{center}
\caption[Broadband contrast for]{Top: Shapes of $DM1$ and $DM2$ in the case of a PIAA unit with a $1$cm pupil. Bottom: Broadband contrast for a hybrid PIAA solution with $1$ cm mirrors. $\Delta \lambda = 120$ nm. The throughput is $0.19$}
\label{FigShapesFor1cm}
\end{figure}

%
%
%
%
%
%
We note that the deformation of the mirrors scales as $D^2/z$ and is a weak function of the choice of pre- and post-apodizers. Namely, for given mirror diameter $D$ and mirror separation $z$, and for the functional form of the pre- and post-apodizers in Eq.~\ref{Eq::PreApod} and Eq.~\ref{Eq::PostApod}, there is a wide range of apodization parameters that will yield deformations below $4$ microns. This means that one can choose a geometry according to the maximal deformation constraint, and then proceed by tuning the apodizers so the broadband contrast constraint is satisfied. Thus, our first step is to satisfy the sag constraint by selecting an adequate geometry, $z = 1$ m in our case. We then proceeded by enforcing the contrast constraint using numerical propagators that evaluate the electric field ringing at M2, which is due to the edge diffraction between M1 and M2. The magnitude of this ringing is the feature that limits the broadband contrast of a given PIAA unit. The purpose of the pre- and post-apodizers is to mitigate these chromatic high frequency edge oscillations that are remapped near the center of M2. These remapped Fresnel rings scale as $\lambda z/ D^2$. Since we chose $\lambda z/ D^2$ to be about 0.01, the apodizers need to be quite strong in order to mitigate for large ringing. Consequently, we chose aggressive apodisers, $\beta = 1.3$ and $\alpha_{M2} = 0.5$, which limits our throughput to $0.19$ and increases our Inner Working Angle (IWA) to $2.4 \lambda/D$.  The loss in throughput is mostly due to the fact that  $DM1$ has to be considerably oversized, and since $\beta> 1$, the actual angular magnification differs from the one for the ideal ray optics designed system: instead of a value close to 2, the angular magnification equals 1.4 along the diagonal. The results are shown Fig.~\ref{FigShapesFor1cm}. The top of Fig.~\ref{FigShapesFor1cm} illustrates the deformation of two $1$ cm DMs separated by $1$ m for the apodisers parameters chosen here. The PSF at three wavelengths across the bandwidth in the visible are shown on the bottom part of Fig.~\ref{FigShapesFor1cm} .

Finally we discuss here the practical implementation of the solution proposed in this subsection. The actual reflective portion of $DM1$ has to be of $1.18$ cm, with a $1$ cm inner region that is responsible for the coronagraphic remapping. In that region the maximal deformation is $4 \; \mu$m, and this region ought to be actuated since it corresponds to the high optical quality part of the beam that will propagate through the coronagraph. The outer edges also exhibit a $4 \; \mu$m deformation, but with a diverging profile and the optical quality of this part of $DM1$ can be crude since this light will not reach the final science camera. $DM2$ exhibits as well a maximal deformation of $4 \; \mu$m. Thus we believe that the solution presented here is a viable alternative to current PIAA solutions, using soon to be available DM technology. This result is of critical importance with respect to the feasibility of a PIAA coronagraph. Because any space based coronagraph will require two sequential DMs for wavefront correction purposes, we are proposing here to accomplish both the coronagraphic suppression and the wavefront correction using the same apparatus. This not only simplifies the design of the instrument, but also eliminates the difficult problem of polishing static PIAA mirrors to sufficient precision. In the design proposed here, since the light remapping surfaces and the actively controllable surfaces are the same, the requirements on the static aberrations (deformation of these optics without any DM command)  are considerably relaxed compared to the case where the remapping optics and the DM are separate. 

\section{Conclusion}
This paper presents recent advances in our theoretical understanding of the impact on contrast of free-space propagation between the two aspherical mirrors designed to generate a Phase Induced Amplitude Apodisation. Our results are two-fold. First, we introduced a semi-analytical tool to compute the diffraction limited field after PIAA coronagraphs of arbitrary geometries. This methodology has already been applied to a number of numerical and experimental studies pertaining to the manufacturability of such coronagraphs and to their performances in the presence of wavefront errors compensated by Deformable Mirrors. We expect this semi-analytical result to be fully integrated in the design of the broadband wavefront control loop in experiments validating the ability of pupil mapping coronagraphs to yield the dynamic range necessary to image exo-earths from space at very close angular separations.

The second conclusion of this paper stems from applying this novel diffractive formalism to design PIAA units on square separable apertures. In the case of exo-earth imaging, it is sensible to explore such geometries since current experimental efforts to integrate together coronagraphs and wavefront control are based on square Deformable Mirrors. We illustrated the performance of several designs over different pupil sizes, and laid down the theoretical foundation for a systematic optimization of such coronagraphs. Finally, we emphasized a class of solutions that is particularly interesting: a PIAA unit composed of two Deformable Mirrors. Indeed, for small pupil sizes, the amplitude remapping necessary to generate an apodisation that produces sufficient contrast for exo-earth imaging can be generated solely with DMs. While the throughput of the designs presented here is reduced to $20 \%$, these solutions considerably simplify manufacturing and wavefront control of starlight suppression systems. We emphasize that the performances of DM based PIAA coronagraphs presented in this communication are illustrative realizations of a vast parameter space. Ongoing theoretical work is being carried out in order to mitigate the loss in performance associated with these solutions. In particular we are currently investigating optimizations that seek to maximize throughput, and therefore angular resolution, for given levels of contrast and bandwidths.

\appendix

\section{Geometrical angular magnification}
We follow the presentation of \cite{2006ApJ...639.1129M} and define the local angular magnification at $M2$ as the spatial frequency of a tilted wavefront at $M1$, as seen from $M2$. Assuming that the incoming wavefront is written as:
\begin{equation}
E_{in}(x,y) = e^{i \frac{2 \pi}{D} \gamma_{M1}(x \cos \phi + y \sin \phi)}
\end{equation}
Then, because of the pupil remapping, the first order of the outgoing wavefront will be:
\begin{equation}
E_{out}(\tilde{x},\tilde{y}) = e^{i \frac{2 \pi}{D} \gamma_{M2}(\tilde{x},\tilde{y})(\tilde{x} \cos \phi + \tilde{y} \sin \phi)}.
\end{equation}
We define the local angular magnification as 
\begin{equation}
\mathcal{M}(\tilde{x},\tilde{y}) = \frac{\gamma_{M2}(\tilde{x},\tilde{y})}{\gamma_{M1}}
\end{equation}
Moreover $E_{out}(\tilde{x},\tilde{y})$ is directly given by the geometrical remapping,
\begin{equation}
E_{out}(\tilde{x},\tilde{y}) = e^{i \frac{2 \pi}{D} \gamma_{M1}(x_{i}(\tilde{x}) \cos \phi + y_{i}(\tilde{y}) \sin \phi)}
\end{equation}
resulting in,
\begin{equation}
\mathcal{M}(\tilde{x},\tilde{y}) = \sqrt{(\frac{\partial g_{1}}{\partial \tilde{x}} \cos \phi)^2 +(\frac{\partial g_{2}}{\partial \tilde{y}} .\sin \phi) ^2}
\end{equation}
Here we define the overall angular magnification using the barycenter of the PSF of an off-axis source. More realistically it should be defined using the maximum of such a PSF, but in practice the two results are very close and the convention chosen in this paper allows us to quickly compute $\mathcal{M}$. Then overall magnification, $\mathcal{M}$, is defined as the average of the local magnification weighted by the pupil transmission function. Carrying out the calculation for the design presented associated to Fig~\ref{Field1DpropPostApodZoom9cm}, for which $\beta<1$,  yields magnifications that oscillate between $2.1$ and $2.2$ depending on $\phi$, the orientation of the off axis source. For simplicity we choose to use the maximum value, $\mathcal{M} = 2.2$

\section{Proof of the analytical propagation theorem}
Our objective is to show that the second order terms in the Taylor expansion of the optical path length about the rays of pupil mapping, namely $G_2(x,y,\tilde{x},\tilde{y})$ in Eq.~\ref{EqTaylorHuygensShort}, reduces to the exponent shown in Eq.~\ref{EqThoremProp}.  For clarity, we begin by repeating this Taylor expansion,
%
\begin{eqnarray}
G_2(x,y,\tilde{x},\tilde{y}) &=&  \frac{1}{2}\frac{\partial^{2} Q}{\partial x^{2}}\big |_{x=x_i,y= y_i} (x-x_i)^{2} \nonumber \\
&+& \frac{\partial^{2} Q}{\partial x \partial y}\big |_{x=x_i, y=_i}(x-x_i)(y-y_i) \nonumber \\
&+& \frac{1}{2}\frac{\partial^{2} Q}{\partial y^{2}}\big |_{x=x_i, y=y_i}(y-y_i)^{2} \label{EqTaylorExp}
\end{eqnarray}
where $x_i = g_1(\tilde{x},\tilde{y})$ and $y_i = g_2(\tilde{x},\tilde{y})$. Note that in the proof which follows, the coordinates $(\tilde{x},\tilde{y})$ appearing in $Q$ are independent of $(x,y)$; the partials are taken only with respect to $x$ and $y$ and then evaluated at the pupil mapped locations $x_i$ and $y_i$.


We proceed by evaluating the three second partial derivatives of the optical path length using Eqs.~\ref{DefOPL} and \ref{DefS}:
\begin{eqnarray}
\frac{\partial ^{2} Q}{\partial x^{2}} &=&  \frac{\partial ^{2} S}{\partial x^{2}} + \frac{\partial ^{2} h}{\partial x^{2}} \nonumber \\
&=&  \frac{\big ( 1+\frac{\partial ^{2} h}{\partial x^{2}}(h-\tilde h)+\frac{\partial h}{\partial x}^{2}   \big ) S(x,y,\tilde{x},\tilde{y})^{2}-\big ( (x-\tilde{x}) + \frac{\partial h}{\partial x}(h-\tilde h)\big )^{2}}{S(x,y,\tilde{x},\tilde{y})^{3}} + \frac{\partial ^{2} h}{\partial x^{2}} \label{firstpartial}
\end{eqnarray}

\begin{eqnarray}
\frac{\partial ^{2} Q}{\partial y^{2}} &=&  \frac{\partial ^{2} S}{\partial y^{2}} + \frac{\partial ^{2} h}{\partial y^{2}} \nonumber \\
&=&  \frac{\big ( 1+\frac{\partial ^{2} h}{\partial y^{2}}(h-\tilde h)+\frac{\partial h}{\partial y}^{2}   \big ) S(x,y,\tilde{x},\tilde{y})^{2}-\big ( (y-\tilde{y}) + \frac{\partial h}{\partial y}(h-\tilde h)\big )^{2}}{S(x,y,\tilde{x},\tilde{y})^{3}} + \frac{\partial ^{2} h}{\partial y^{2}} \label{secondpartial}
\end{eqnarray}

\begin{eqnarray}
\frac{\partial ^{2} Q}{\partial x \partial y} &=&  \frac{\partial ^{2} S}{\partial x \partial y} + \frac{\partial ^{2} h}{\partial x \partial y} \nonumber  \\
&=&  \frac{\big ( \frac{\partial h}{\partial x} \frac{\partial h}{\partial y} +(h- \tilde h)\frac{\partial ^{2} h}{\partial x \partial y} \big ) S(x,y,\tilde{x},\tilde{y})^{2}}{S(x,y,\tilde{x},\tilde{y})^{3}} \nonumber \\ &-& \frac{\big ( (x-\tilde x) +(h-\tilde h)\frac{\partial h}{\partial x} \big )\big ( (y-\tilde y) +(h-\tilde h)\frac{\partial h}{\partial y} \big )}{S(x,y,\tilde{x},\tilde{y})^{3}} + \frac{\partial ^{2} h}{\partial x \partial y} \label{thirdpartial}
\end{eqnarray}

 Because of symmetries, we need only carry out the derivation for the second derivative with respect to $x$ and the mixed derivative. We re-arrange $\frac{\partial ^{2} Q}{\partial x^{2}}$ and $\frac{\partial ^{2} Q}{\partial x \partial y}$ as:
\begin{equation}
\frac{\partial ^{2} Q}{\partial x^{2}} = \frac{1}{S} + \frac{\partial ^{2} h}{\partial x^{2}} (1+\frac{h - \tilde{h}}{S} )+\frac{ (S \frac{\partial h}{\partial x})^2-\big((x-\tilde{x}) + \frac{\partial h}{\partial x}(h - \tilde{h}) \big)^2}{S^3}
\label{Eq::SecondDevGen}
\end{equation}

\begin{equation}
\frac{\partial ^{2} Q}{\partial x \partial y} =  \frac{\partial ^{2} h}{\partial x \partial y}  (1+\frac{h - \tilde{h}}{S} )+\frac{ S^2\frac{\partial h}{\partial x}\frac{\partial h}{\partial y}- \big((x-\tilde{x}) + \frac{\partial h}{\partial x}(h - \tilde{h}) \big) \big((y-\tilde{y}) + \frac{\partial h}{\partial y}(h - \tilde{h}) \big)}{S^3}
\label{Eq::HybridDevGen}
\end{equation}
Note that  we have not yet replaced the coordinates by the actual ray optics values; thus Eqs.~\ref{Eq::SecondDevGen} and \ref{Eq::HybridDevGen} are valid for all  $(x,y)$ and $(\tilde{x},\tilde{y})$.

Next we rewrite Eqs. 15 and 16, which give the height $h$ of $M1$ as a function of $(x,y)$ and the pupil mapped locations $f_1(x,y)$ and $f_2(x,y)$:
\begin{eqnarray}
\frac{\partial h}{\partial x} \left (1+\frac{h(x,y) -\tilde{h}(f_1(x,y),f_2(x,y))}{S(x,y,f_1(x,y),f_2(x,y))} \right ) &=&\frac{f_1(x,y) - x}{S(x,y,f_1(x,y),f_2(x,y))}\label{EqFirstDevQuotx}\\
\frac{\partial h}{\partial y} \left (1+\frac{h(x,y) -\tilde{h}(f_1(x,y),f_2(x,y))}{S(x,y,f_1(x,y),f_2(x,y))} \right ) &=&\frac{f_2(x,y) - y}{S(x,y,f_1(x,y),f_2(x,y))}\label{EqFirstDevQuoty}
\end{eqnarray}
or, 
\begin{eqnarray}
S(x,y,f_1,f_2) \frac{\partial h}{\partial x}  = -(h(x,y) - \tilde{h}(f_1,f_2)) \frac{\partial h}{\partial x}  - (x - f_1)\label{EqFirstDevQuotxHSimple}\\
S(x,y,f_1,f_2) \frac{\partial h}{\partial y}  = -(h(x,y) - \tilde{h}(f_1,f_2)) \frac{\partial h}{\partial y}  - (y - f_2)\label{EqFirstDevQuotyHSimple}.
\end{eqnarray}
%
%

We now use Eqs.~\ref{EqFirstDevQuotxHSimple} and \ref{EqFirstDevQuotyHSimple} to replace the partial derivatives of $h$ in Eqs.~\ref{Eq::SecondDevGen} and \ref{Eq::HybridDevGen}.  Beginning with Eq.~\ref{Eq::SecondDevGen}, we see that it is composed of three additive terms.  Evaluating at $(x_i,y_i)$, the third one can be simplified as follows:
\begin{eqnarray}
&&\frac{ (S \frac{\partial h}{\partial x})^2-\big((x-\tilde{x}) + \frac{\partial h}{\partial x}(h - \tilde{h}) \big)^2}{S^3} \Big |_{x_i,y_i} \\
&=& \frac{ (S(x_i,y_i,\tilde{x},\tilde{y}) \frac{\partial h}{\partial x}|_{x_i,y_i})^2-(S(x_i,y_i,\tilde{x},\tilde{y}) \frac{\partial h}{\partial x}|_{x_i,y_i})^2}{S(x_i,y_i,\tilde{x},\tilde{y})^3}. \\
&=& 0
\end{eqnarray}
where we have used Eq.~\ref{EqFirstDevQuotxHSimple}  evaluated at $x_i = g_1(\tilde{x},\tilde{y})$ and $y_i = g_2(\tilde{x},\tilde{y})$ to cancel the two terms in the numerator. In order to simplify the second term of Eq.~\ref{Eq::SecondDevGen} we re-write Eq.~\ref{EqFirstDevQuotxHSimple} as:
\begin{equation}
\frac{\partial h}{\partial x}  (S(x,y,f_1,f_2) +(h(x,y) - \tilde{h}(f_1,f_2))) = \frac{\partial h}{\partial x} Q(x,y,f_1,f_2) = (f_1 - x).  \label{EqDevHandQBis}
\end{equation}
Taking the partial derivative of Eq.~\ref{EqDevHandQBis} with respect to $x$ yields:
\begin{equation}
  \frac{\partial ^{2} h}{\partial x^{2}} Q(x,y,f_1,f_2) + \frac{\partial h}{\partial x} \frac{\partial Q}{\partial x} = \dpart{f_1}{x} -1
\end{equation}
We now evaluate this equation at $x_i = g_1(\tilde{x},\tilde{y})$ and $y_i = g_2(\tilde{x},\tilde{y})$. By definition, the ray optics coordinates are such that $\frac{\partial Q}{\partial x} \Big |_{x_i,y_i} = 0$. Consequently,
\begin{equation}
  \frac{\partial ^{2} h}{\partial x^{2}} \Big |_{(x_i,y_i)} Q(x_i,y_i,f_1,f_2) = \dpart{f_1}{x} \Big |_{(x_i,y_i)} -1
\end{equation}
or, dividing through by $S$,
\begin{equation}
  \frac{\partial ^{2} h}{\partial x^{2}} (1+\frac{h - \tilde{h}}{S} )\Big |_{(x_i,y_i)} = \frac{1}{S_0} ( \dpart{f_1}{x} \Big |_{(x_i,y_i)} -1)
  \end{equation}
where we define $S_0$ as
\begin{equation}
S_0  = S(g_1(\tilde{x},\tilde{y}),g_2(\tilde{x},\tilde{y}),\tilde{x},\tilde{y}) = S(x_i,y_i,f_1(x_i,y_i),f_2(x_i,y_i)) .
\end{equation}
Substituting into Eq.~\ref{Eq::SecondDevGen} leaves
\begin{equation}
\frac{\partial ^{2} Q}{\partial x^{2}}  \Big |_{x_i,y_i} = \frac{1}{S_0} +  \frac{1}{S_0} ( \dpart{f_1}{x} \Big |_{x_i,y_i} -1) =   \frac{1}{S_0} \dpart{f_1}{x}  \Big |_{x_i,y_i}.
\end{equation}

Turning to Eq.~\ref{Eq::HybridDevGen}, we see that it is composed of two additive terms. The second one can be simplified as follows:
\begin{eqnarray}
&&\frac{ S^2 \frac{\partial h}{\partial x} \frac{\partial h}{\partial y} -(x-\tilde{x} + \frac{\partial h}{\partial x}(h - \tilde{h})) (y-\tilde{y} + \frac{\partial h}{\partial y}(h - \tilde{h})) } {S^3} \Big |_{x_i,y_i} \nonumber \\
&=& \frac{ S(x_i,y_i,\tilde{x},\tilde{y})^2 \frac{\partial h}{\partial x} \Big |_{x_i,y_i} \frac{\partial h}{\partial y} \Big |_{x_i,y_i} - S(x_i,y_i,\tilde{x},\tilde{y})^2 \frac{\partial h}{\partial x} \Big |_{x_i,y_i}  \frac{\partial h}{\partial y} \Big  |_{x_i,y_i} }{S(x_i,y_i,\tilde{x},\tilde{y})^3} \nonumber \\
&=& 0
\end{eqnarray}
where we have used Eq.~\ref{EqFirstDevQuotxHSimple} and Eq.~\ref{EqFirstDevQuotyHSimple}  evaluated at $x_i = g_1(\tilde{x},\tilde{y})$ and $y_i = g_2(\tilde{x},\tilde{y})$ to cancel the two terms in the numerator. In order to simplify the first term of Eq.~\ref{Eq::HybridDevGen} we again use Eq.~\ref{EqDevHandQBis}.
%
%
Taking the partial derivative with respect to $y$ yields:
\begin{equation}
  \frac{\partial ^{2} h}{\partial x \partial y } Q(x,y,f_1,f_2) + \frac{\partial h}{\partial x} \frac{\partial Q}{\partial y} = \dpart{f_1}{y} .
\end{equation}
We now evaluate this equation at $x_i = g_1(\tilde{x},\tilde{y})$ and $y_i = g_2(\tilde{x},\tilde{y})$. By definition, the ray optics coordinates are such that $\frac{\partial Q}{\partial y} \Big |_{x_i,y_i} = 0$.  Consequently, after again dividing by $S$,
\begin{equation}
 \frac{\partial ^{2} h}{\partial x \partial y } (1+\frac{h(x,y) - \tilde{h}(f_1,f_2)}{S(x,y,f_1,f_2)} )\Big |_{x_i,y_i} = \frac{1}{S_0}  \dpart{f_1}{y} \Big |_{x_i,y_i} .
\end{equation}

Note that we also could have also started with Eq.~\ref{EqFirstDevQuotxHSimple} and taken its partial derivative as a function of $x$ and found
\begin{equation}
 \frac{\partial ^{2} h}{\partial x \partial y}  (S(x,y,f_1,f_2) + h(x,y) -\tilde{h}(f_1,f_2)) \Big |_{x_i,y_i}  = \dpart{f_2}{x}  |_{x_i,y_i} 
\end{equation}
Since the condition $\dpart{f_2}{x} \Big |_{x_i,y_i}  = \dpart{f_1}{y} \Big |_{x_i,y_i} $ has to be enforced in order for continuous curvature solutions for $M1$ and $M2$ to exist, these two results are equivalent.

We now substitute these results into Eq.~\ref{Eq::HybridDevGen} to find:
\begin{equation}
\frac{\partial ^{2} Q}{\partial x \partial y}  \Big |_{x_i,y_i} =   \frac{1}{S_0}  \dpart{f_1}{y} \Big |_{x_i,y_i}  =   \frac{1}{S_0}  \dpart{f_2}{x} \Big |_{x_i,y_i}.
\end{equation}
This finishes our proof since the $\frac{\partial ^{2} OPL}{\partial y^{2}} $ can be calculated using exactly the same procedure as $\frac{\partial ^{2} OPL}{\partial x^{2}} $.

\section{Semi-analytical propagation of the pre-apodiser through the remapping mirrors}

In this appendix we detail how we propagate the pre-apodiser using the special functions $ FresnelC$ and $ FresnelS$. For clarity we work in one dimension, the full two dimensional result can be obtain using a tensor product in the case of separable designs. Our goal is to derive an expression for
\begin{equation}
E_{OutPre}(\tilde{x}) = \frac{1}{2 i \lambda Z} \int_{-(1+\alpha_{M1})D/2}^{(1+\alpha_{M1})D/2}E_{Pre}(x)  e^{\frac{ i \pi }{\lambda S_0 A_X(\tilde{x})^2}  (x-x_i(\tilde{x})^{2} } dx
\end{equation}
that only involves function evaluations of $ FresnelC$ and $ FresnelS$.  We achieve this by choosing a pre-apodiser with the functional form in Eq.~\ref{Eq::PreApod},
\begin{equation}
A_{pre}(x)  = \left\{\begin{array}{clc}
\frac{1}{2} \left(1 +\ cos [ \frac{2 \pi}{D \alpha_{M1}} (x+ \frac{D}{2}) ] \right)  & \mbox{for } & - \frac{ \alpha_{M1} D}{2}< x  < \frac{D}{2}  \\
1 & \mbox{for} & -\frac{D}{2} < x < \frac{D}{2}   \\
\frac{1}{2} \left(1 + \cos [ \frac{2 \pi}{D \alpha_{M1}} (x- \frac{D}{2}) ]\right)  & \mbox{for} & \frac{D}{2}< x  < \frac{ \alpha_{M1} D}{2}\end{array}  \right.
\end{equation}
Using superposition we can write that the field at $M2$ resulting from the propagation of such a pre-apodiser through a pupil mapping unit is the sum of the contribution of the right, left, and center portions of this such an apodisation function,
\begin{equation}
E_{OutPre}(\tilde{x}) = E_{Left}(\tilde{x}) + E_{Center}(\tilde{x}) + E_{Right}(\tilde{x}).
\end{equation}
We adopt the following shorthanded notation for the Fresnel integrals
\begin{equation}
\mathcal{F}_{\nu}(x) = \int_{\nu (x-1/2)}^{\nu (x+1/2)} e^{i \frac{\pi}{2} u^2} du.
\end{equation}
The contribution of the center portion of the apodiser can be computed easily using the results described in this paper
\begin{equation}
E_{Center}(\tilde{x})  = \mathcal{F}_{\sqrt{\frac{2}{\lambda z}} \frac{D}{A_{X}(\tilde{x})}}(\frac{x_{i}(\tilde{x})}{D}).
\end{equation}
To computed the contribution of the left side of the pre-apodiser we use the Euler formula 
\begin{equation}
\cos [ \frac{2 \pi}{D \alpha_{M1}} (x- \frac{D}{2}) ] = \frac{1}{2} ( e^{i\frac{2 \pi}{D \alpha_{M1}} (x- \frac{D}{2})  } + e^{- i \frac{2 \pi}{D \alpha_{M1}} (x- \frac{D}{2}) }).
\end{equation}
For each exponential component we add the exponents of the pre-apodiser and the propagation kernel and we complete the squares. We then factor out of the integral the exponential terms that only depend on $\tilde{x}$ and re-write it as a function of $\mathcal{F}_{\nu}(x)$. We do not detail here the actual derivation and only provide the reader with the final result of such a calculation,
\begin{eqnarray}
E_{Left}(\tilde{x}) &=& \mathcal{F}_{\sqrt{\frac{2}{\lambda z}} \frac{\alpha_{M1} D }{2 A_{X}(\tilde{x})}}(\frac{x_{i}(\tilde{x})}{D} + \frac{1}{2} + \frac{\alpha_{M1}}{4}) \\
&+&  e^{i (\frac{\pi}{\alpha_{M1}} + \frac{2 \pi}{\alpha_{M1} D} x_{i}(\tilde{x})  - \frac{4 \pi \alpha_{M1} \lambda z A_{X}(\tilde{x})}{D^2})}\mathcal{F}_{\sqrt{\frac{2}{\lambda z}} \frac{\alpha_{M1} D }{2 A_{X}(\tilde{x})}}(\frac{x_{i}(\tilde{x})}{D} + \frac{1}{2} + \frac{\alpha_{M1}}{4} - \frac{2 \alpha_{M1} A_{X}(\tilde{x}) \lambda z}{D^2}) \nonumber \\
&+& e^{i (- \frac{\pi}{\alpha_{M1}} - \frac{2 \pi}{\alpha_{M1} D} x_{i}(\tilde{x})  - \frac{4 \pi \alpha_{M1} \lambda z A_{X}(\tilde{x})}{D^2})}\mathcal{F}_{\sqrt{\frac{2}{\lambda z}} \frac{\alpha_{M1} D }{2 A_{X}(\tilde{x})}}(\frac{x_{i}(\tilde{x})}{D} + \frac{1}{2} + \frac{\alpha_{M1}}{4} + \frac{2 \alpha_{M1} A_{X}(\tilde{x}) \lambda z}{D^2}). \nonumber
\end{eqnarray}
Similarly we find that
\begin{eqnarray}
E_{Right}(\tilde{x}) &=& \mathcal{F}_{\sqrt{\frac{2}{\lambda z}} \frac{\alpha_{M1} D }{2 A_{X}(\tilde{x})}}(\frac{x_{i}(\tilde{x})}{D} - \frac{1}{2} - \frac{\alpha_{M1}}{4}) \\
&+&  e^{i (-\frac{\pi}{\alpha_{M1}} + \frac{2 \pi}{\alpha_{M1} D} x_{i}(\tilde{x})  - \frac{4 \pi \alpha_{M1} \lambda z A_{X}(\tilde{x})}{D^2})}\mathcal{F}_{\sqrt{\frac{2}{\lambda z}} \frac{\alpha_{M1} D }{2 A_{X}(\tilde{x})}}(\frac{x_{i}(\tilde{x})}{D} - \frac{1}{2} - \frac{\alpha_{M1}}{4} - \frac{2 \alpha_{M1} A_{X}(\tilde{x}) \lambda z}{D^2}) \nonumber \\
&+& e^{i (+\frac{\pi}{\alpha_{M1}} - \frac{2 \pi}{\alpha_{M1} D} x_{i}(\tilde{x})  - \frac{4 \pi \alpha_{M1} \lambda z A_{X}(\tilde{x})}{D^2})}\mathcal{F}_{\sqrt{\frac{2}{\lambda z}} \frac{\alpha_{M1} D }{2 A_{X}(\tilde{x})}}(\frac{x_{i}(\tilde{x})}{D} - \frac{1}{2} - \frac{\alpha_{M1}}{4} + \frac{2 \alpha_{M1} A_{X}(\tilde{x}) \lambda z}{D^2}). \nonumber
\end{eqnarray}
As a consequence the propagation of a pre-apodiser that has the functional form prescribed by Eq.~\ref{Eq::PreApod} through the pupil mapping unit can be computed using solely functional evaluation of the Fresnel special functions.


\clearpage

\clearpage

\clearpage

\clearpage

\clearpage

\clearpage

\clearpage

\clearpage

\clearpage

\begin{figure}
\begin{center}
\includegraphics[width=4.5in]{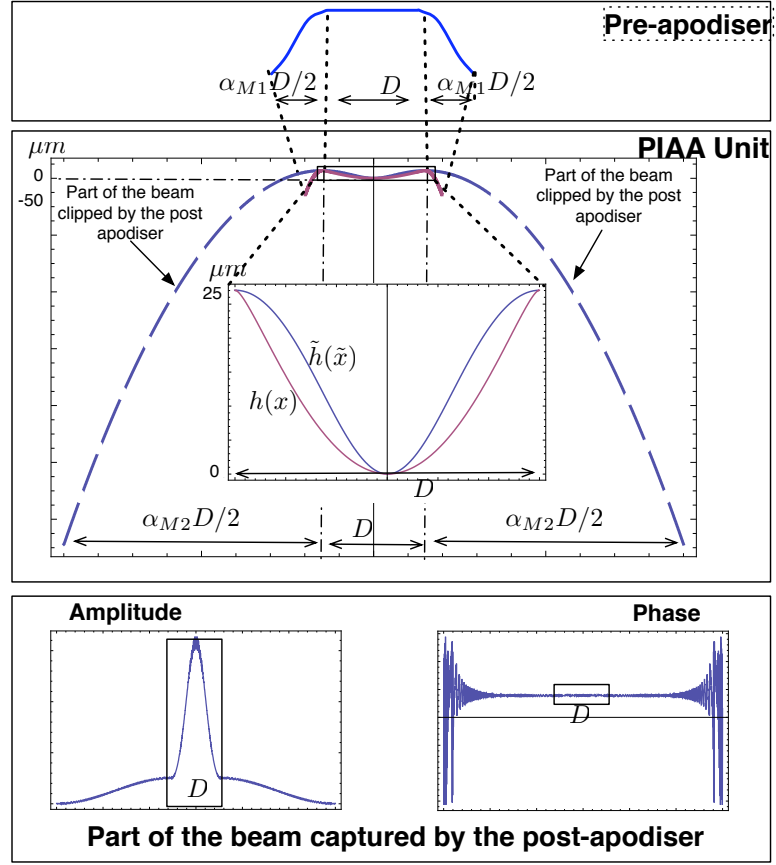}
\end{center}
\caption[Actual values of the PIAA mirror shapes]{Mirror profiles in the presence of pre and post apodisers.  The top and bottom panels are similar to the previous figure. Middle panel: the central beam profile is obtained using the area preserving mirror shapes calculated according to the ray optics model and using the saturated mirror induced apodisation in Eq.~\ref{Eq::PostApod}. The outer portion of $M1$ is chosen to have a constant negative curvature. The focal length of this outer parabolic portion is chosen such that the curvature of $M1$ remains constant. This results in a greatly expanded area over which the edge oscillations are located in the plane of $M2$. The bulk of these edge oscillations is then clipped by the post-apodiser. Note that the parabolic outer region of $M2$ does not need to be actually manufactured since it does not contribute to the field seen by the coronagraph.}
\label{CartonnRealValuesShapes}
\end{figure}

\clearpage

\clearpage

\clearpage

\clearpage

\clearpage

\clearpage

\begin{table}[htbf]
\begin{center}
\begin{tabular}{c|c|c|c}Physical size of $M1$ & Size of the useful region of $M1$ & Physical size of $M2$ & Throughput \\\hline $(1+\alpha_{M1}) D$ & $D$ & $D$ & $1/(1 + \alpha_{M1})^2$ \\\hline $D$ & $(1 - \alpha_{M1} ) D$ & $ (1 - \alpha_{M2}) D$ & $(1 - \alpha_{M2})^2$\end{tabular}
\end{center}
\caption{Two equivalent conventions to describe hybrid apodised PIAA designs. Note that to first order they are equivalent in terms of throughput. In this paper we chose the one in the top row}
\label{Tab::PIAAConv}
\end{table}

\clearpage

\begin{table}
\begin{tabular}{c|c|c|c|c}
Pupil Size & Maximal Deformation & Throughput & Angular magnification & IWA \\\hline \hline  $9$ cm  & $585\;  \mu$m  & $0.89$ & $2.2$ & $1.8$ \\ $3$ cm & $62 \; \mu$m & $0.55$ & $2.1$ & $1.9$ \\  $1$ cm  & $4 \; \mu$m  & $0.19$ & $1.4$ & $2.3$
\end{tabular}
\caption{Summary of the performances of the designs discussed above. The separation of the mirrors is maintained constant at $1$ m and the contrast is constrained to be below $10^{-10}$ along the diagonal within a $20\%$ bandwidth around $600$ nm}
\label{Tab::PIAAPerfs}
\end{table}

\end{document}